\documentclass[preprint,preprintnumbers,amsmath,amssymb,showkeys]{revtex4-1}

\usepackage{graphicx}
\usepackage{amsmath}

\usepackage{amsmath,graphicx}
\usepackage{amsthm}
\usepackage{amsfonts}
\usepackage{graphicx}
\usepackage{multirow}
\usepackage{subcaption}
\usepackage{float}
\usepackage{verbatim}
\usepackage{amsfonts}
\usepackage{amssymb}
\usepackage{dcolumn}
\usepackage{color}

\usepackage{lipsum}
\usepackage{array}
\usepackage[table,dvipsnames]{xcolor}

\usepackage[english]{babel}

\begin{document}

\title{{ Entropic characterization of Tunneling and State Pairing in a Quasi-Exactly Solvable Sextic Potential}}

\author{Angelina N. Mendoza Tavera }%
\email{anmtavera@gmail.com}
\affiliation{Departamento de F\'{i}sica, Universidad Aut\'onoma Metropolitana Unidad Iztapalapa, San Rafael Atlixco 186, 09340 Cd. Mx., M\'exico}

\author{Adrian M. Escobar Ruiz}%
\email{admau@xanum.uam.mx}
\affiliation{Departamento de F\'{i}sica, Universidad Aut\'onoma Metropolitana Unidad Iztapalapa, San Rafael Atlixco 186, 09340 Cd. Mx., M\'exico}

\author{Robin P. Sagar}%
\email{sagar@xanum.uam.mx}
\affiliation{Departamento de Química, Universidad Aut\'onoma Metropolitana Unidad Iztapalapa, San Rafael Atlixco 186, 09340 Cd. Mx., M\'exico}

\begin{abstract}
We analyze the (de)localization properties of a quasi-exactly solvable (QES) sextic potential $
V_{\text{QES}}(x) = \tfrac{1}{2}(x^6 + 2x^4 - 2(2\lambda + 1)x^2),
$ as a function of the tunable parameter $\lambda \in [-\tfrac{3}{4}, 6]$. For $\lambda > -\tfrac{1}{2}$, the potential exhibits a symmetric double-well structure, with tunneling emerging for the ground state level at $\lambda \approx 0.732953$. {For the lowest energy states \( n = 0,1,2,3 \), we construct physically meaningful variational wavefunctions that $i)$ respect parity symmetry under the transformation \( x \rightarrow -x \), $ii)$ exhibit the correct asymptotic behavior at large distances, and $iii)$ allow for exact analytical Fourier transforms. 
Variational energies match Lagrange Mesh and available exact analytical QES results with relative errors $\simeq10^{-8}$ for \(n = 0, 1, 2\) and \(\simeq 10^{-6}\) for the third excited state $n=3$. We demonstrate that entropic measures (Shannon entropy, Kullback–Leibler, and Cumulative Residual Jeffreys divergences) surpass conventional variance-based methods in revealing tunneling transitions, wavefunction symmetry breaking, and quantum state pairing. Our results confirm that the Beckner–Białynicki-Birula–Mycielski entropic uncertainty relation holds across all examined values of $n$ and $\lambda$. The quality of the trial density functions is also validated by the small $\sim10^{-10}$ Cumulative Residual Jeffreys divergences from the available exact QES solutions. }

\end{abstract}

\maketitle

\section{Introduction}

{
In quantum mechanics, quasi-exactly solvable (QES) systems constitute a distinguished class of spectral problems of the form ${\cal H}\,\psi = E \,\psi$, for which a finite portion of the energy spectrum and the corresponding eigenfunctions can be obtained exactly and analytically, while the remainder of the spectrum lacks closed-form solutions. QES systems thus lie between exactly solvable models, like the harmonic oscillator or hydrogen atom, and fully non-solvable systems that demand purely numerical methods.

A systematic and unifying framework for quasi-exact solvability is provided by algebraic methods, particularly those involving hidden Lie algebraic structures~\cite{Turbiner1988QuasiexactlysolvablePA, TURBINER1987181, U1988exact}. In such cases, the Hamiltonian can be recast as a quadratic combination (with constant coefficients) of the generators of a finite-dimensional representation of a Lie algebra, most commonly $\mathfrak{sl}(2,\mathbb{R})$. This representation-theoretic perspective allows for the construction of a finite-dimensional invariant subspace of the Hilbert space, often spanned by polynomials of bounded degree. Within this subspace, the Hamiltonian acts exactly, yielding closed-form eigenvalues and eigenfunctions. As such, the analysis of quantum systems governed by QES potentials offers a unique opportunity to explore nontrivial quantum phenomena within a partially analytic framework.

For the QES sextic potential, the existence of this hidden algebraic structure has significant implications for the dynamical and spectral properties of the system. One remarkable feature is the appearance of energy-reflection (ER) symmetry, i.e. the spectrum is symmetric under $E \rightarrow -E$ after a shift, a discrete property not evident in the original formulation, but which emerges naturally from the algebraic construction~\cite{PhysRevA.59.1791}. Another key feature is the existence of a generating function for a set of orthogonal polynomials $P_n(E)$ in the energy variable $E$, which encode the solvable part of the spectrum and are connected to the underlying differential operators and recurrence relations~\cite{10.1063/1.531373}.

Moreover, the sextic QES potential is of particular interest due to its rich structural behavior, including the emergence of a double-well configuration as a function of a tunable parameter. As this parameter is varied, the potential undergoes a qualitative transition, leading to the penetration of certain quantized energy levels into classically forbidden regions. This enables quantitative analysis of tunneling effects and spontaneous symmetry breaking.

In general, the double-well potential plays a fundamental role in theoretical and computational chemistry, offering deep insights into a broad range of physical phenomena. Characterized by two minima separated by a potential barrier, this simple model captures the essential physics of bistable configurations. In chemistry, double-well potentials are particularly important for understanding molecular isomerization, proton transfer reactions, and quantum tunneling processes, where particles transition between two nearly equivalent states. In quantum computing, the double-well potential is employed as a model for qubits in quantum dots or Josephson junction systems. Similarly, in quantum optics, it is used to simulate the behavior of Bose--Einstein condensates confined in optical double-well potentials~\cite{Entropytunneling2015,Bose-Einstein,qubits}.

From a quantum mechanical perspective, the double-well potential serves as a prototypical system for exploring symmetry breaking, energy level splitting, and the subtle interplay between potential energy landscapes and wavefunction behavior. It also provides a natural setting for the study of semiclassical approximations and quantum tunneling effects, which are crucial for accurately describing reaction rates at low temperatures.

In this context, information-theoretic tools (particularly Shannon entropy in both position and momentum space) provide a sensitive and quantitative means of characterizing the evolution of a system's quantum states~\cite{PhysRevLett.99.263601, PANOS2019384, NASCIMENTO2018401, Royqua.25596, Sun201300089, SONG20151402,Jiaoqua.25375, ChakladarPhysRevA.110.042819, VMajerník1996, Laguna201400156, SAGAR200872}. These entropic measures capture subtle changes in wavefunction localization and delocalization, offering insight into the competing influences of confinement and kinetic uncertainty. Position-space entropy reflects the spatial extent of the probability distribution, while momentum-space entropy quantifies the corresponding dynamical uncertainty, and together they encode the underlying phase-space structure of the quantum state.

An information-theoretic formulation of the uncertainty principle~\cite{Bialynicki-Birula:1975igq, WilliamB1975, PhysRevA.50.3065, Hertz_2019,RevModPhys.89.015002} offers a powerful alternative to the standard deviation-based Heisenberg inequality~\cite{PhysRev.34.163}. In this framework, quantum uncertainty is quantified using the Shannon entropies \( S_x \) and \( S_p \), associated with the probability distributions of position and momentum, respectively. The entropic uncertainty relation, known as the Beckner–Białynicki-Birula–Mycielski (BBM) entropic uncertainty relation \cite{Bialynicki-Birula:1975igq, WilliamB1975},
asserts that the sum \( S_x + S_p \) is bounded from below by a universal constant, independent of the specific quantum state. In one spatial dimension, this bound is given by \( S_x + S_p \geq 1+\log(\pi \, \hbar) \). Unlike the Heisenberg inequality, which may become ill-defined for states with divergent variances, the entropic bound remains valid and informative for a broader class of states, including those exhibiting strong localization or long tails. As such, entropic uncertainty relations provide a more general and often sharper characterization of the fundamental limits imposed by quantum mechanics on simultaneous knowledge of conjugate observables.

Numerous double-well systems have naturally been examined from an information-theoretic perspective. Ammonia (\(\mathrm{NH}_{3}\)), for example, has served as a prototype, with its tunneling dynamics analyzed through various entropic measures~\cite{tunnelingammonia}. In one dimension, the Shannon entropy of a hyperbolic double-well potential has been studied as a function of well depth and width; interestingly, although exact polynomial solutions arise for discrete parameter sets, neither an underlying hidden algebra nor a possible quasi-exactly-solvable (QES) character has yet been clarified~\cite{Shannonhyperbolicdoublewell}. Other work has probed how stochastic fluctuations in the height and width of a symmetric barrier modify both the semiclassical transmission probability and the associated Shannon entropy~\cite{stochasticdoublewell}. Most recently, researchers have explored the entropic information content of fractional Schrödinger systems with symmetric hyperbolic double wells, focusing on its dependence on the well depth and on the fractional order that characterizes the kinetic operator~\cite{fractionaldoublewell}.

Importantly, the availability of exact QES eigenfunctions facilitates rigorous benchmarking of numerical entropy computations, enabling a systematic investigation of entropy-based signatures associated with the emergence of the double-well regime. In doing so, this work fills a key gap in the literature by offering a tunable, analytically tractable reference model for entropic analyses of quantum tunneling and localization. Moreover, QES models provide ideal benchmarks for validating numerical algorithms in quantum information theory, including applications in machine learning, entropic diagnostics, and variational quantum algorithms.

Therefore, this study aims to elucidate the informational aspects of quantum state evolution in QES systems and to demonstrate the utility of entropy as a diagnostic tool for quantum transitions driven by potential deformation. The primary objective of the present work is to explore the localization--delocalization behavior of a sextic QES system using entropic measures across three distinct regimes of the parameter $\lambda$: (i) for negative values $\lambda \leq -\frac{1}{2}$, where the QES potential exhibits a single-well structure; (ii) for intermediate values $-\frac{1}{2} < \lambda < \lambda_{c}^{n=0} \approx 0.732953$, where a double-well structure begins to form, yet without observable tunneling effects; and (iii) for $\lambda > \lambda_{c}^{n=0}$, where tunneling phenomena emerge in the ground state. Here, $\lambda_{c}^{n=0} \approx 0.732953$ denotes the critical value at which the ground-state wavefunction first begins to tunnel across the central barrier. As $\lambda$ increases beyond this point, the excited states begin to penetrate the classically forbidden region.

A central component of the analysis is the comparison between the standard deviations $\Delta x$ and $\Delta p$ and the corresponding Shannon entropies in position and momentum space. By contrasting the Heisenberg uncertainty relation with the entropic uncertainty sum {BBM}, this approach offers complementary insights into the quantum state's spread and uncertainty beyond second-moment descriptions.

The paper is organized as follows. In Section~\ref{s2}, we introduce the quasi-exactly solvable sextic potential and describe its key features as a tunable model interpolating between single- and double-well configurations. Section~\ref{s3} presents the variational and numerical methods used to obtain high-accuracy eigenfunctions and energies, emphasizing the reliability of the results through comparison with the Lagrange mesh method and available exact QES solutions. In Section~\ref{s4}, we explore the spatial and spectral structure of the probability densities, focusing on the emergence of energy level pairing and tunneling signatures as the parameter $\lambda$ increases. Section~\ref{s5} focuses on the Shannon entropy in both position and momentum representations and analyzes how these quantities reveal structural transitions not captured by standard uncertainty relations. Afterwards, in Section~\ref{s6}, the tunneling and pairing phenomena are investigated by means of Kullback--Leibler and the Cumulative Residual Jeffreys divergences. The end of this section provides a Table summarizing the main findings of our study. The paper concludes in Section~\ref{s7} with final remarks, and directions for future research, including the novelty of studying a continuously tunable family of potentials and the advantages of entropic diagnostics in identifying quantum phase-like transitions.

} 

\section{Model Definition and Setup}
\label{s2}
We consider the following one-dimensional spectral problem in non-relativistic quantum mechanics:
\begin{equation}
\label{Hp}
    {\cal H}\,\psi(x)\ = \ E\,\psi(x) \ ,   \qquad \quad \psi(x) \in {\cal L}^2 \ , 
\end{equation}
defined on the real line, $x \in (-\infty,\infty)$. The corresponding Hamiltonian operator for our specific problem is of the form:
\begin{equation}
\label{HQES}
    {\cal H} \ = \ -\frac{\hbar^2}{2\,m}\frac{d^2}{dx^2} \ + \ V^{\rm QES}(x;\lambda) \ , 
\end{equation}
where $m$ denotes the mass of the particle, and
\begin{equation}
\label{vqes}
V^{\rm QES}(x;\lambda)\ = \  \frac{1}{2}\,(\,x^{6}\ + \ 2\,x^{4}\ - \ 2\,(2\,\lambda+1)\, \ x^{2}\,) \ ,
\end{equation}
is the quasi-exactly solvable sextic potential with $\lambda$ a real parameter. 
Hereafter, we will adopt atomic units $\hbar=1$, $m=1$.

\begin{figure}[h]
\begin{center}
\includegraphics[width=12cm]{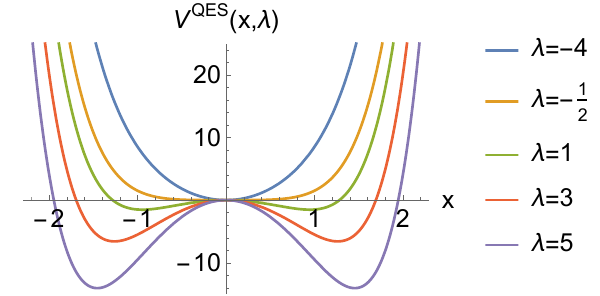}
\caption{\small QES confining sextic potential $V^{\rm QES}(x;\lambda)$ in (\ref{vqes}) for different values of the parameter $\lambda$. For $\lambda>-\frac{1}{2}$ it develops two symmetric degenerate minima, and the maximum located at $x = 0$ corresponds to a potential energy $V = 0$.
}
\label{F1}
\end{center}
\end{figure}
For \( \lambda \) arbitrary, the QES potential admits infinitely many bound states and no scattering states are present, see Fig.~\ref{F1}. The spectrum is therefore entirely discrete, which is a notable feature of this model.

For certain \( \lambda \), a particularly interesting phenomenon occurs: the potential develops a double-well structure. Specifically, when \( \lambda = -\frac{1}{2} \), the potential~(\ref{vqes}) acquires two symmetric, degenerate minima. In this regime, tunneling effects can arise, leading to instanton-like corrections to the semiclassical spectrum. In the large-\( \lambda \) limit, the positions of the minima in (\ref{vqes}) tend to 
\[
x_\pm \ \approx \ \pm\sqrt{2} \left(\frac{\lambda}{3}\right)^{1/4} \, ,
\]
and the corresponding potential values are approximately given by
\[
V^{\rm QES}(x_\pm) \ \approx \ \frac{4}{9} \lambda \left(3  \,-\, 2\, \sqrt{3\, \lambda} \right) \, .
\]
Analytical estimation shows that the width of the potential well scales as \( \lambda^{1/4} \), while its depth scales more rapidly as \( \lambda^{3/2} \). Consequently, the ratio of width to depth decreases asymptotically as \( \lambda^{-5/4} \), indicating that in the large-\( \lambda \) limit, the well becomes sharply confining and strongly localized around its minima.

The Hamiltonian~(\ref{HQES}) possesses parity symmetry under the transformation \( x \rightarrow -x \), which ensures that the eigenstates alternate between even (symmetric) and odd (antisymmetric) parity. More recently, it has been shown that energy-reflection (ER) symmetry~\cite{PhysRevA.59.1791} is also a structural feature of the model~\cite{GAlvarez, Kreshchuk_2019}.

Importantly, the Hamiltonian~(\ref{HQES}) is characterized by a hidden \( \mathfrak{sl}_2(\mathbb{R}) \) Lie algebra symmetry~\cite{Turbiner1988QuasiexactlysolvablePA}. When \( \lambda  \) is a non-negative (semi)integer $\lambda=\frac{2N+1}{2},N$, \( N \in \mathbb{Z}^+ \), this algebra admits a finite-dimensional irreducible representation. In such cases, the system allows for an explicit algebraic construction of solely \( (N+1) \) exact (odd)even-parity eigenfunctions and their associated eigenvalues.

Further connections of this system with nonperturbative quantum phenomena have been explored in~\cite{Kozçaz}. There, it is shown that for integer values of \( \lambda \), the exact cancellation between real and complex nonperturbative saddle contributions in the semiclassical expansion can be understood within the instanton-calculus framework, illustrating the deep interplay between algebraic solvability and path-integral formulations.

Finally, an interesting limit is obtained when \( \lambda = -j - \frac{1}{2} \), with \( j \) a positive integer. In this case, the potential reduces to the well-known sextic \( \mathcal{PT} \)-symmetric quasi-exactly solvable potential introduced in~\cite{Bender2005}, which further enriches the landscape of QES systems by incorporating concepts from non-Hermitian quantum mechanics and symmetry-breaking phenomena.

\section{Energy spectrum and wavefunctions}
\label{s3}

For the lowest states $n=0,1,2,3$ and different values of the parameter $\lambda \in [-\frac{3}{4},6]$, we obtain approximate solutions (spectrum and wavefunctions) for the spectral problem (\ref{Hp}) using the variational method.

\subsection{Variational results}

At fixed $\lambda$, the variational trial function of the $n$th-state is chosen to be:

\begin{equation}
\label{trialf}
    \psi_{n}(x;\lambda) \ = \ Q_n(x;\lambda)\,e^{-\frac{x^4}{4}} \ ,
\end{equation}

\noindent where the polynomial factor $Q_n=Q_n(x;\lambda)$ depends on the parity symmetry. For even $n=0,2$, it reads

\begin{equation}
    Q_{n}(x;\lambda) \ = \ 1 \ + \ \sum_{j=1}^{k_{\rm even}} a_j\,x^{2j} \ ,
\end{equation}
whereas for odd $n=1,3$, we have
\begin{equation}
    Q_{n}(x;\lambda) \ = \ x \ + \ \sum_{j=1}^{k_{\rm odd}} b_j\,x^{2j+1} \ ,
\end{equation}
here $a_j,\,b_j$ are linear variational parameters. In this way, the so-constructed trial functions have the properties: 

\begin{itemize}
    \item They incorporate the exact parity symmetry $(x \rightarrow -x)$ explicitly. 
    \item The dominant asymptotic behavior of the phase in the exponential decay factor, \( e^{-x^4/4} \), namely \( \sim \frac{x^4}{4} \) as \( |x| \to \infty \), is exactly reproduced.

    \item An appealing feature of these trial functions is that their Fourier transforms can be computed analytically, which is instrumental for analyzing the delocalization in momentum space.
\end{itemize}

\noindent For $\lambda$ a positive integer number, the finite number of $(\lambda+1)$ exact QES eigenfunctions appears in the form \cite{Turbiner1988QuasiexactlysolvablePA}
\[
\psi_N^{\rm QES}(x) \ = \ P_N(x) \,e^{-x^4/4 - x^2/2},
\]
where $P_N(x)$ is a polynomial of degree $N=0,1,2\,\ldots,\lambda$, c.f. (\ref{trialf}). The value $N=0$ corresponds to the ground state $(n=0)$, $N=1$ to the second excited state ($n=2$), $N=2$ to the fourth excited state ($n=4$) and so on. The QES solutions do not possess exact analytical Fourier transforms.

As a first step, for each fixed $\lambda$, some of the variational parameters $a_j,\,b_j$ are determined by imposing the orthogonality condition $\langle \psi_n \mid \psi_{\tilde n} \rangle = \delta_{n\,\tilde n}$, which is trivially satisfied between states of opposite parity (i.e., even and odd states). Subsequently, the remaining parameters are obtained by minimizing the energy functional.

On the other hand, for any fixed value of $\lambda$, highly accurate numerical eigenvalues can be efficiently computed using the \textit{LagrangeMesh Mathematica Package} (LMMP)~\cite{JCR}. Owing to the method’s exceptional efficiency and its precise control over numerical accuracy, reliable results are attainable even with limited computational resources, such as a standard personal laptop.

The values of the integers \(k_{\rm even}\) and \(k_{\rm odd}\), which determine the number of terms in the trial function~(\ref{trialf}), are optimally chosen to ensure a relative error of approximately \(10^{-8}\) for the lowest energy states (\(n = 0, 1, 2\)) and \(10^{-6}\) for the third excited state (\(n = 3\)), when comparing the variational energies to the corresponding numerical results obtained using the Lagrange-Mesh method, for all values of \(\lambda\) considered. {The optimal values of the variational parameters $a_i,b_j$ in (\ref{trialf}) corresponding to the lowest states \( n = 0, 1, 2, 3 \), computed for several values of the coupling constant \( \lambda \), are presented in Appendix~\ref{Ap1}.
}

\subsection{Energy spectrum}

\begin{figure}[!th]
    \includegraphics[scale=0.9]{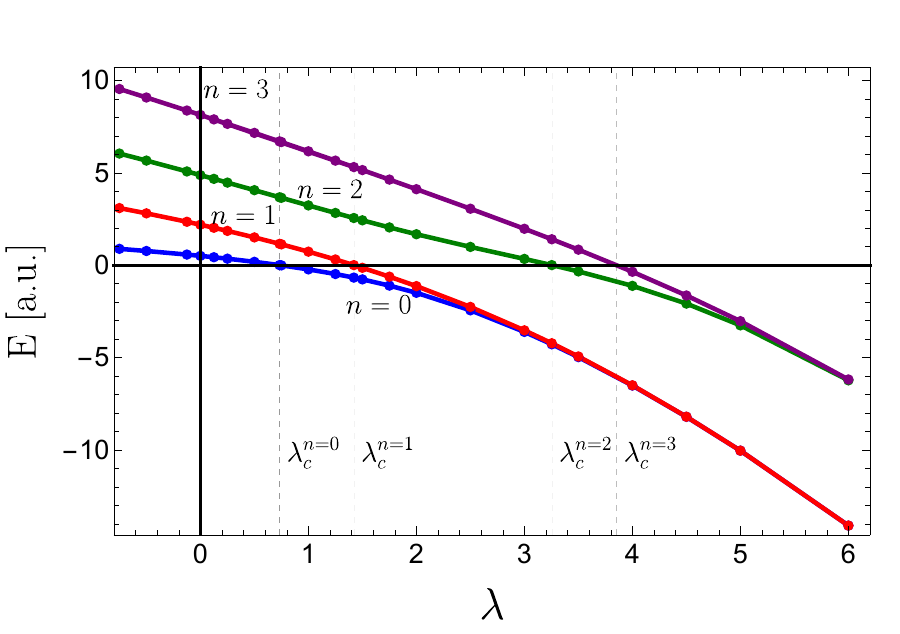}
    \caption{\small \textbf{QES potential \( V^{\rm QES} \):} Energy levels \( E \) vs.\ coupling \( \lambda \) for the lowest states \( n = 0, 1, 2, 3 \). The line \( E = 0 \) corresponds to the central barrier peak for \( \lambda > -\frac{1}{2} \), and to the potential minimum for \( \lambda < -\frac{1}{2} \) (see Fig.~\ref{F1}). Vertical dashed lines indicate critical couplings \( \lambda_c^n \) where \( E_n = 0 \), signaling tunneling onset. The line at \( \lambda = 0 \) separates algebraic and non-algebraic sectors.
}
    \label{ES}
\end{figure}

In Fig. \ref{ES}, we show the variational energy $E_n$ of the lowest states $n=0,1,2,3$ as a function of the parameter $\lambda$. The curves $ E_n = E_n(\lambda) $ are smooth, monotonically decreasing functions. The phenomenon of level pairing becomes clearly evident in the region $ E < 0 $, where the energy levels enter the classically forbidden domain. This generic behavior is characteristic of the symmetric double-well potential, where tunneling leads to the formation of nearly degenerate energy levels and instanton contributions \cite{primer_aniceto_2019, multiinstantons_zinnjustin_2004, exact_kamata_2024}. In the sextic QES system, each $n$th-level crosses into the forbidden region at a specific critical value of the parameter $\lambda= \lambda_c^{n}$, marking the onset of this pairing structure.

As the energy levels approach and cross the barrier at \( E = 0 \), corresponding to the maximum of the double-well potential, a characteristic pairing of levels begins to emerge, indicative of the tunneling-induced quasi-degeneracy typical in such systems. In the regime \( \lambda < 0 \), no exact analytical solutions are known within the infinite tower of bound states. However, for positive integer values \( \lambda = N \), the system admits \( N + 1 \) exact solutions, a signature feature of the sextic QES model. Moreover, we define critical coupling values \( \lambda_c^{n} \) as the points in $\lambda-$space at which the \( n \)-th energy level crosses the \( E = 0 \) line, marking the transition into the classically forbidden region. These critical values exhibit a strict hierarchy, \( \lambda_c^{n} < \lambda_c^{n+1} \), reflecting the increasing coupling strength required for higher excited states to become bound. It is important to note that varying \( \lambda \) alters not only the spectrum but also the shape of the underlying potential; thus, the energy evolution depicted here corresponds to a one-parameter family of sextic potentials, rather than a single fixed potential landscape. {A leading-order asymptotic analysis to estimate the energy $E$ in the limits \( \lambda \to \pm\infty \) is presented in Appendix~\ref{Ap2}.
}

\section{Probability Densities}
\label{s4}

Here, for the states \( n = 0, 1, 2, 3 \), we employ the optimal trial wavefunctions~(\ref{trialf}) together with their corresponding Fourier transforms to obtain the probability densities in both position and momentum space.

\vspace{-0.2cm}

\subsection{Position space}

\begin{figure}[h]
\subfloat[Ground state $n=0$]{\includegraphics[scale = 0.42]{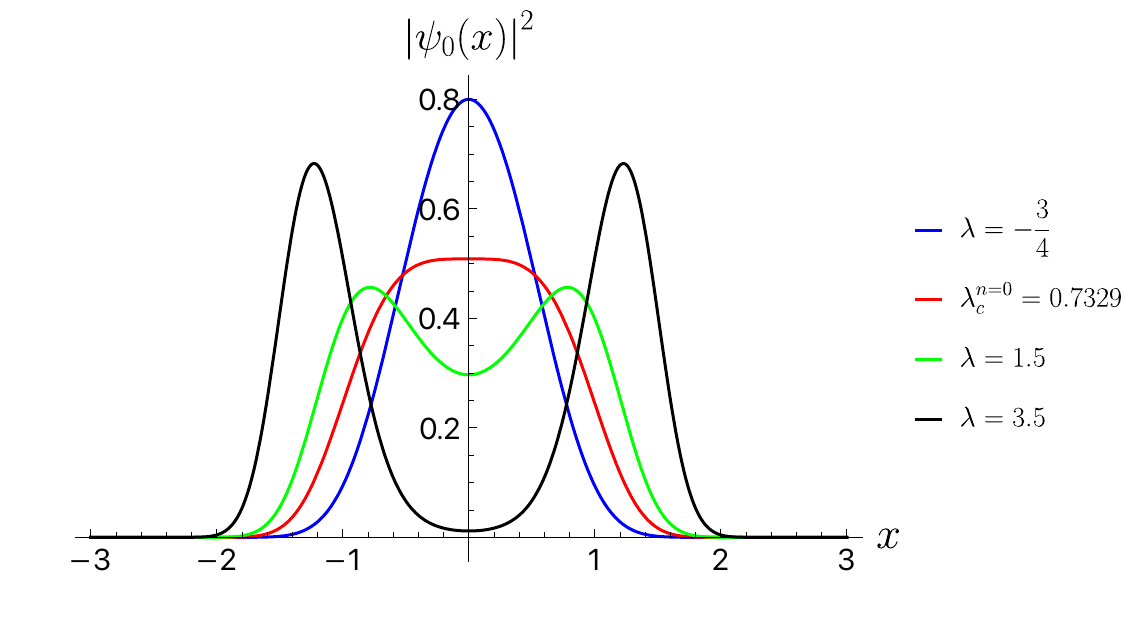}}
\hspace{0.1cm}
\subfloat[First excited state $n=1$]{\includegraphics[scale = 0.42]{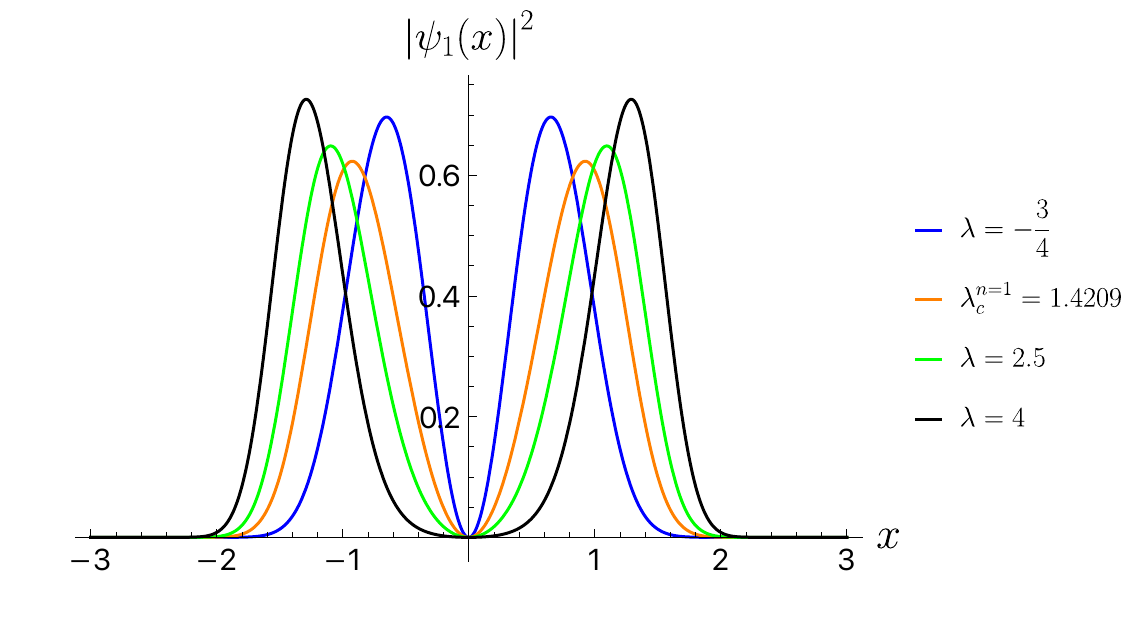}}
\hspace{0.1cm}
\subfloat[Second excited state $n=2$]{\includegraphics[scale = 0.42]{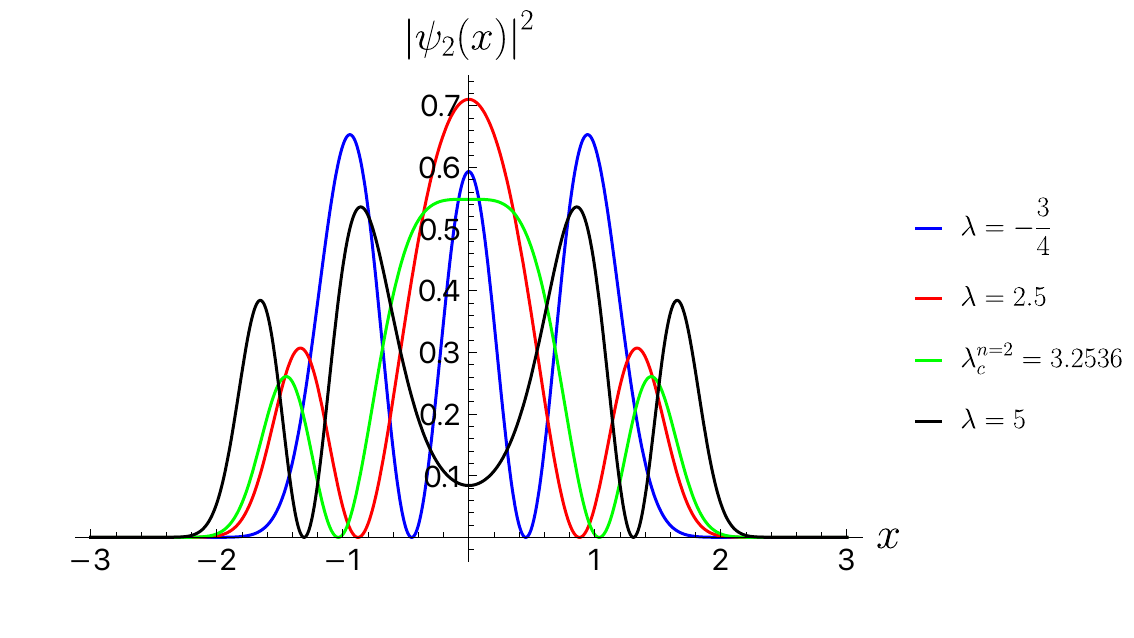}} 
\hspace{0.1cm}
\subfloat[Third excited state $n=3$]{\includegraphics[scale = 0.42]{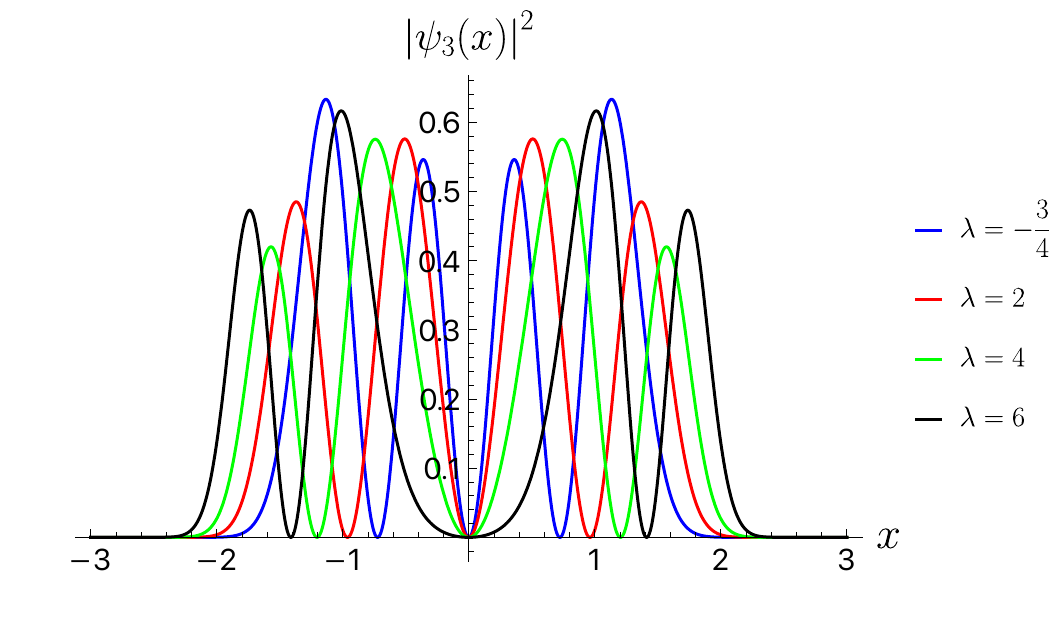}} 
\caption{\small Probability density $|\psi_n(x)|^2$ for the states $n=0,1,2,3$ and different $\lambda$. For each fixed \( n \), the critical value \( \lambda_c^{n} \) denotes the coupling strength at which the \( n \)-th energy level vanishes, i.e., \( E_n = 0 \). This value marks the onset of tunneling for the \( n \)-th state. As \( \lambda \) increases beyond this threshold, the wavefunctions become more localized in the outer wells. For even-parity states (\( n = 0, 2 \)) a pronounced transformation, developing a central minimum and becoming strongly localized in the outer wells, appears. }
\label{psipos}
\end{figure}

Figure~\ref{psipos} presents the probability densities $|\psi_n(x)|^2$ of the lowest four eigenstates ($n = 0, 1, 2, 3$) of the QES sextic potential as functions of position $x$, for various values of the parameter $\lambda$. Each panel corresponds to a fixed quantum number $n$ and illustrates how the spatial profiles of the wavefunctions evolve as $\lambda$ increases.

Although the QES potential develops a double-well structure for $\lambda \gtrsim -\tfrac{1}{2}$, the wells are initially too shallow to support significant tunneling. As $\lambda$ increases and the well depth grows, tunneling behavior appears once the energy of a given state $n$ falls below the central barrier. This occurs when $\lambda$ exceeds the state-dependent critical coupling $\lambda_c^{n}$.

A notable feature in the tunneling regime is the emergence of \textit{quasi-degenerate level pairing}, where adjacent energy levels (e.g., $n = 0, 1$, $n = 2, 3$ and so on) approach one another in energy. This near-degeneracy arises due to exponentially suppressed tunneling splittings between symmetric and antisymmetric combinations of states localized in opposite wells. Although the levels remain non-degenerate, the splitting can become too small to resolve numerically or visually.

For instance, panel~(a) in Figure~\ref{psipos} displays the ground state ($n=0$), which transitions from a single-peaked distribution at low values of $\lambda$ to a double-peaked profile for $\lambda > \lambda_{c}^{n=0} \approx0.7329$, signaling the onset of wavefunction delocalization across both wells. A similar transition from 3 to 4 peaks is observed for the second excited state, shown in panel~(c).

Although the odd-parity excited states, shown in panels~(b) and~(d), also exhibit changes in the shape of their corresponding probability densities, these transformations are less abrupt than those observed in the even-parity states~(a) and~(c). This behavior arises because the lower-energy symmetric state of each pair undergoes a more pronounced reshaping as $\lambda$ increases, adapting its profile to more closely resemble that of the higher-energy antisymmetric state in the pair. Importantly, as $\lambda$ increases, the symmetric (off-center) nodes of the excited states gradually shift outward, toward the minima of the potential wells. This outward movement reflects the progressive deepening of the wells and the growing localization of the wavefunction lobes around them. The central node in odd-parity states (e.g., $n = 1$ and $n = 3$) remains fixed at $x = 0$ due to symmetry, while the outer nodes respond sensitively to the shape of the potential landscape.

Overall, Figure~\ref{psipos} captures the crossover from single-well to double-well-dominated behavior in the QES sextic potential. The evolution of the probability density profiles, including the outward shift of symmetric nodes and the appearance of near-degenerate doublets, provides clear evidence of the onset of tunneling and quantum interference between the two wells, governed by the critical couplings $\lambda_c^{(n)}$.

\begin{figure}[ht]
    \centering

    \begin{subfigure}{0.45\textwidth}
        \centering
        \includegraphics[width=\textwidth]{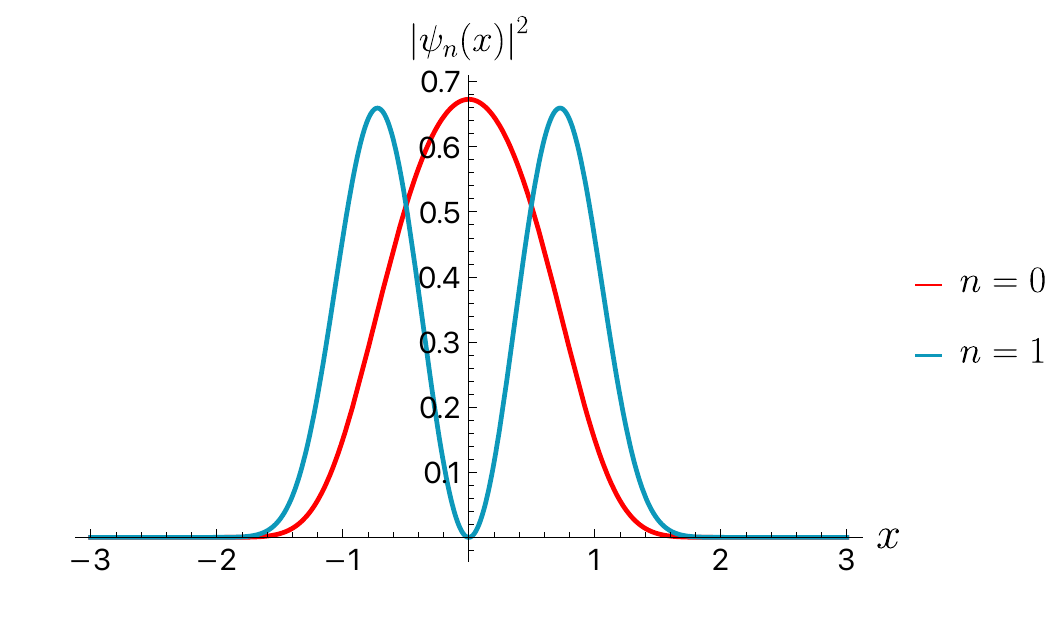}
        \caption{$\lambda = 0$, low-lying states}
    \end{subfigure}\hfill
    \begin{subfigure}{0.45\textwidth}
        \centering
        \includegraphics[width=\textwidth]{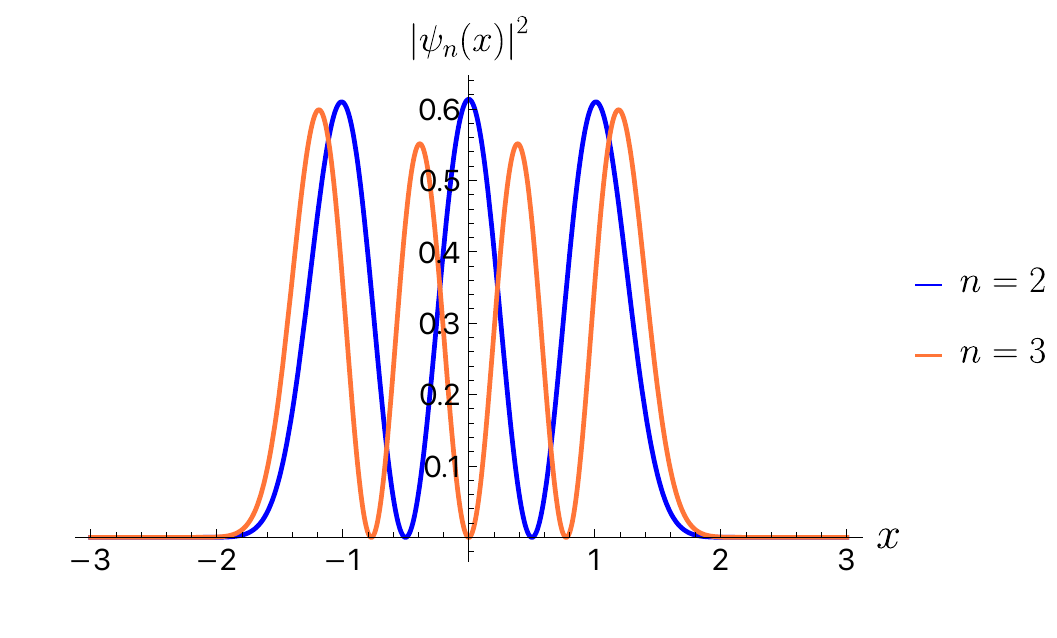}
        \caption{$\lambda = 0$, higher excited states}
    \end{subfigure}

    \vspace{0.4cm}

    \begin{subfigure}{0.45\textwidth}
        \centering
        \includegraphics[width=\textwidth]{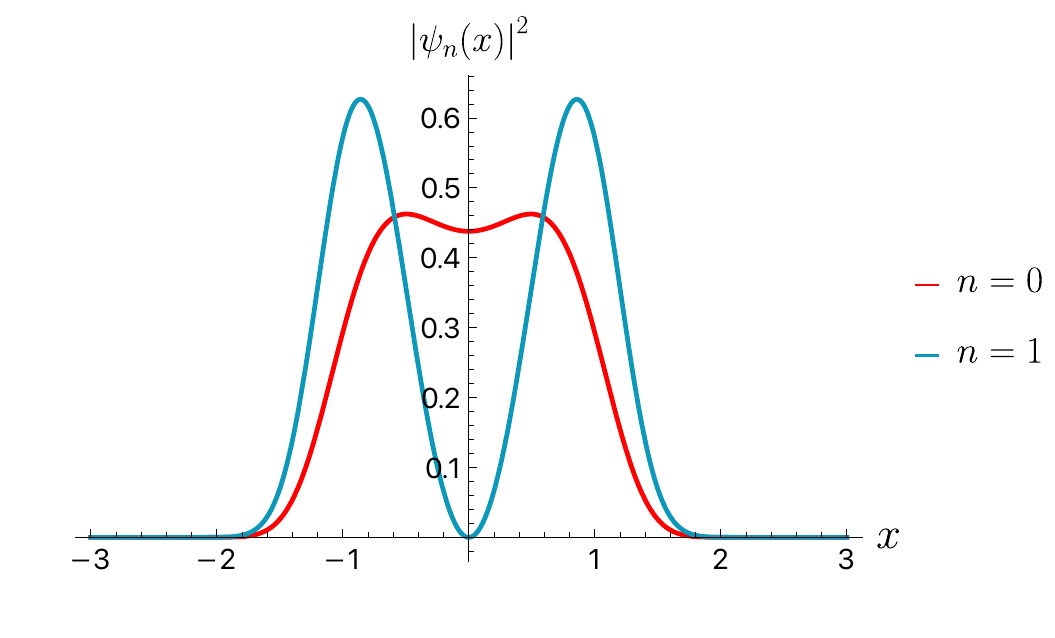}
        \caption{$\lambda = 1$, low-lying states}
    \end{subfigure}\hfill
    \begin{subfigure}{0.45\textwidth}
        \centering
        \includegraphics[width=\textwidth]{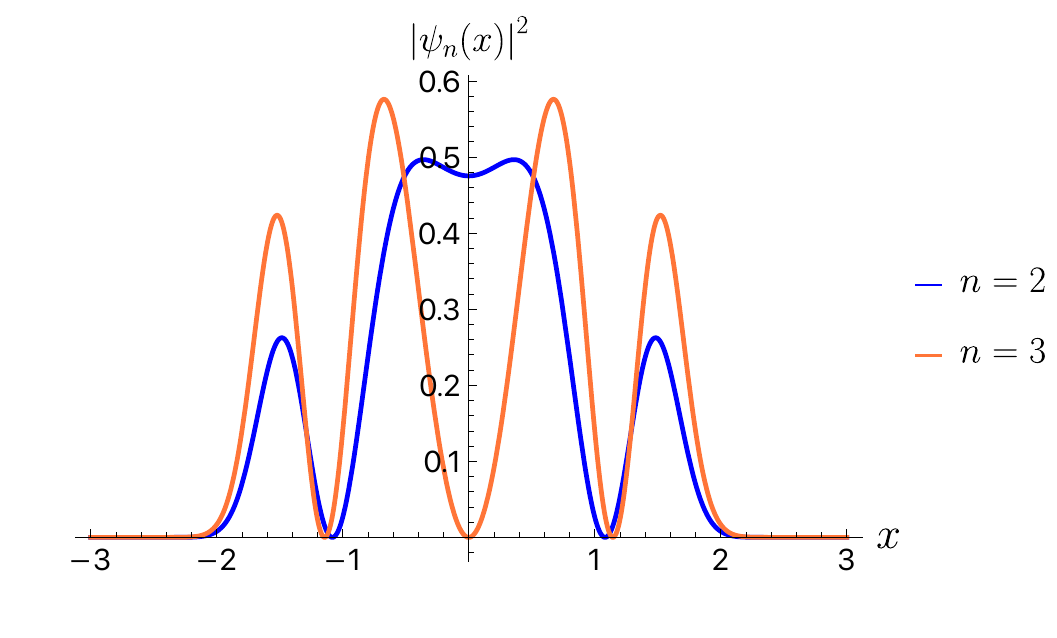}
        \caption{$\lambda = 3.5$, higher excited states}
    \end{subfigure}

    \vspace{0.4cm}

    \begin{subfigure}{0.45\textwidth}
        \centering
        \includegraphics[width=\textwidth]{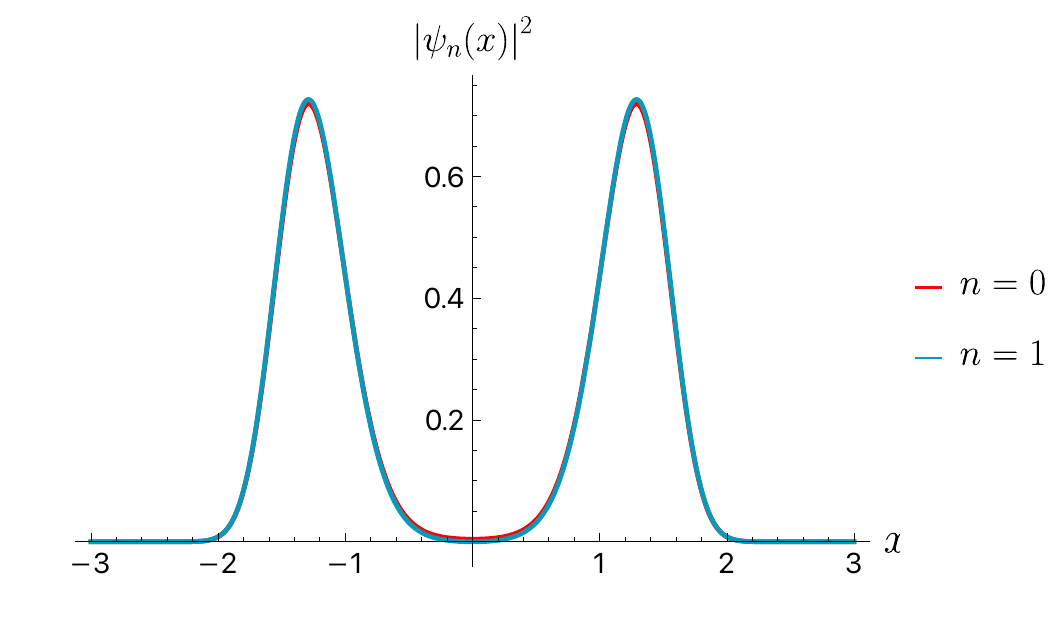}
        \caption{$\lambda = 4$, low-lying states}
    \end{subfigure}\hfill
    \begin{subfigure}{0.45\textwidth}
        \centering
        \includegraphics[width=\textwidth]{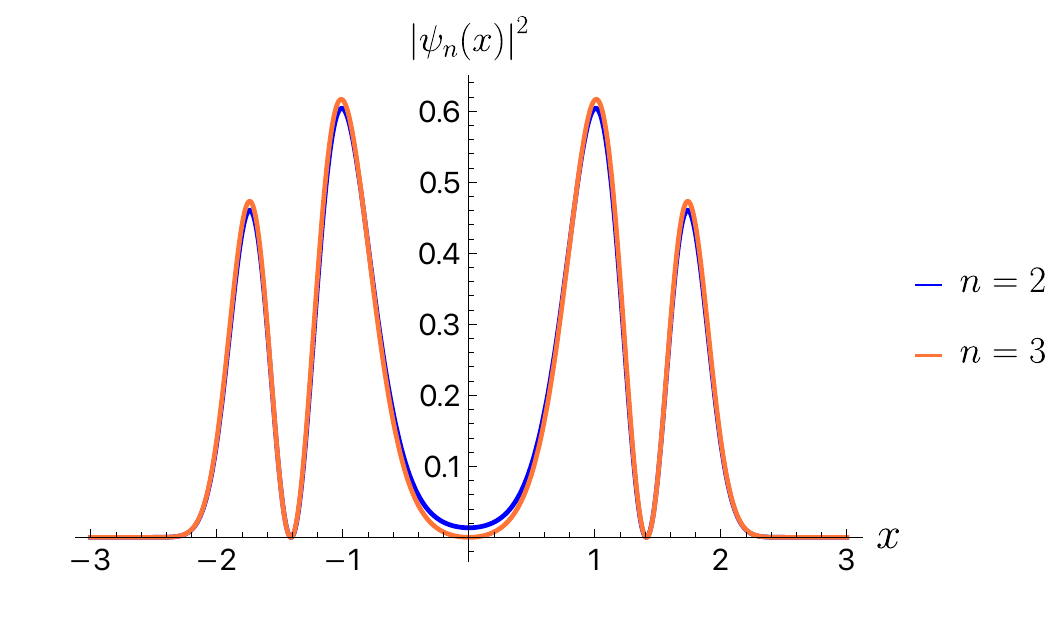}
        \caption{$\lambda = 6$, higher excited states}
    \end{subfigure}
    \caption{\small Probability densities $|\psi_n(x)|^2$ for selected paired states ($n = 0, 1$ and $n = 2, 3$) at fixed coupling values $\lambda = 0$ (top row), $\lambda = 1 \quad\text{and} \quad 3.5$ (middle row), and $\lambda = 4 \quad\text{and} \quad 6$ (bottom row). 
    Left column: low-lying states; right column: higher excited states.}
    \label{fig:six-panels}
\end{figure}

\clearpage

Figure~\ref{fig:six-panels} illustrates the probability densities $|\psi_n(x)|^2$ for the two-lower pairs $n = 0, 1$ and $n=2, 3$ of the QES  sextic potential, evaluated at several representative values of the parameter: $\lambda = 0$, $\lambda = 1$, $\lambda = 3.5$, $\lambda = 4$, and $\lambda = 6$. Each row corresponds to a fixed $\lambda$, with the left column displaying the low-lying states ($n = 0, 1$) and the right column showing the higher excited states ($n = 2, 3$). This layout reveals the progressive emergence of tunneling and quasi-degenerate level pairing as the central barrier becomes increasingly pronounced with growing $\lambda$.

At weak coupling ($\lambda = 0$, top row), the potential is shallow and nearly parabolic. The eigenstates exhibit distinct nodal structures: $\psi_0(x)$ peaks at the origin, $\psi_1(x)$ has a single node at $x = 0$, and the higher excited states $\psi_2(x)$ and $\psi_3(x)$ show more oscillatory behavior, with non-negligible amplitude near the center. These features reflect the absence of a significant barrier and the dominance of single-well confinement.

At intermediate coupling ($\lambda = 1$, second row), the potential begins to form a double-well structure. The ground state $\psi_0(x)$ develops a visible dip at the origin—a \textit{quasi-node}—suggesting incipient tunneling across the central barrier. Meanwhile, $\psi_1(x)$ retains its node at $x = 0$ but becomes increasingly localized near the developing well minima. {In contrast, the wavefunctions $\psi_2(x)$ and $\psi_3(x)$ remain largely unaffected at this value of $\lambda$, as these states still lie well above the classical barrier and are therefore not significantly influenced by the potential.} The symmetric off-center nodes of all wavefunctions shift outward, indicating the separation of classical turning points and the formation of a forbidden region around the origin.

At stronger coupling ($\lambda = 3.5$ and $\lambda = 4$), the double-well is fully developed, and the wavefunctions exhibit clear signatures of tunneling-induced \textit{level pairing}. The states $\psi_0(x)$ and $\psi_1(x)$ become nearly indistinguishable in spatial extent, differing primarily by parity. {Similarly, $\psi_2(x)$ and $\psi_3(x)$ begin to exhibit pairing behavior, as for 
$\lambda=3.5$, the state $\psi_2(x)$ has already surpassed its critical value $\lambda_{c}^{n=2} \approx 3.2536$, and tunneling effects start to alter the shape of its probability density.} In this regime, the central barrier strongly suppresses probability density at $x = 0$ in all even-parity states, and the energy splitting within each pair becomes exponentially small, dominated by instanton-like tunneling contributions.

At very strong coupling ($\lambda = 6$, bottom right panel), the effect is even more dramatic: the higher excited states $\psi_2(x)$ and $\psi_3(x)$ are nearly identical in envelope and peak positions, with only their parity distinguishing them. The suppression of the wavefunction at the origin becomes greatly pronounced in the symmetric states $n=0,2$, and the localization within the individual wells is nearly total. This behavior signifies that tunneling between wells is now extremely rare, and the paired energy levels are effectively quasi-degenerate. We note that, in the present study, we do not present the wavefunctions themselves but rather their associated probability densities $|\psi^2(x)|$ (and $|\phi^2(p)|$), since these are the relevant quantities that enter into the informational entropic analysis.

Therefore, Figure~\ref{fig:six-panels} provides another clear depiction of the transition from single-well to double-well confinement as a function of $\lambda$. It shows how wavefunction deformation, node dynamics, and parity symmetry conspire to generate quasi-degenerate level pairs through tunneling processes, culminating in exponentially localized states at large coupling.

\begin{figure}[ht]
    \centering
    \begin{minipage}{0.45\textwidth}
        \includegraphics[width=\textwidth]{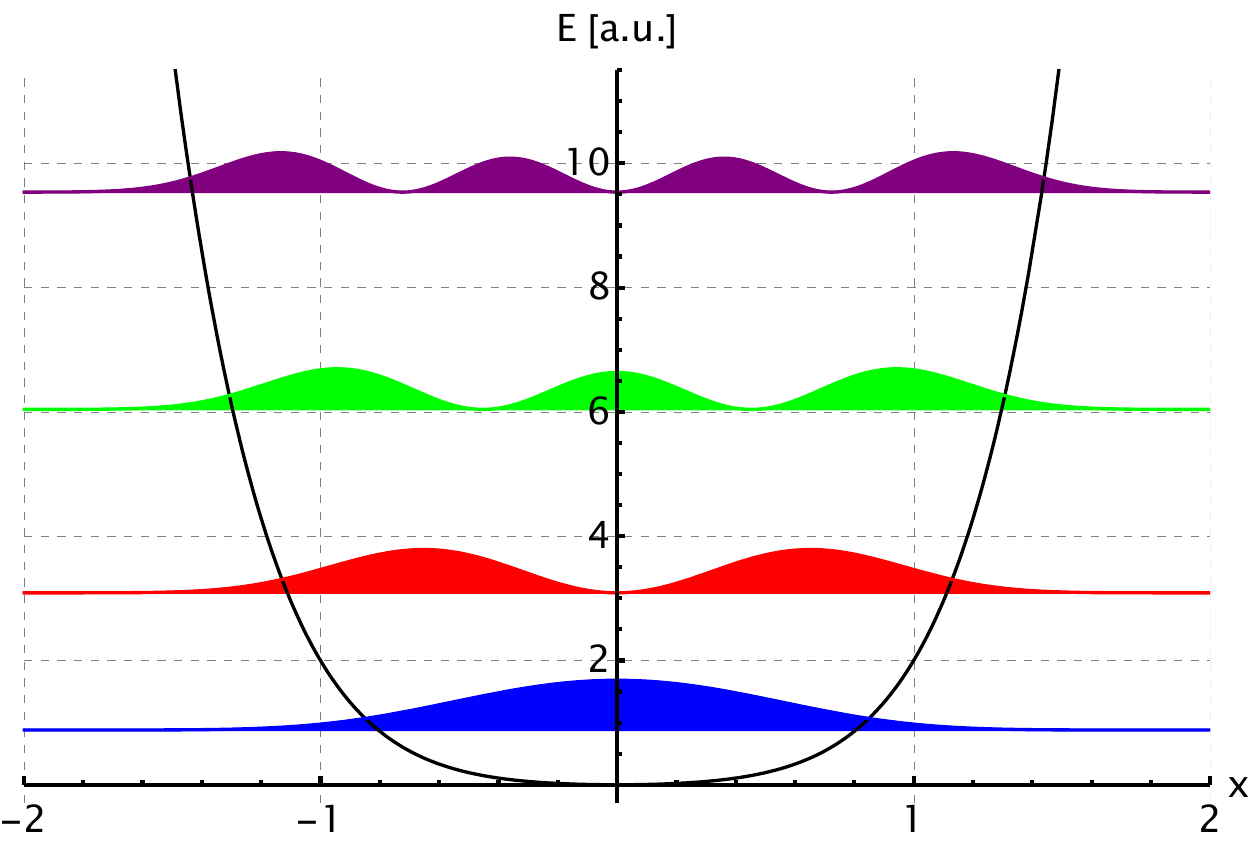}
        \caption*{(a) $\lambda=-3/4$}
    \end{minipage}\hfill
    \begin{minipage}{0.46\textwidth}\includegraphics[width=\textwidth]{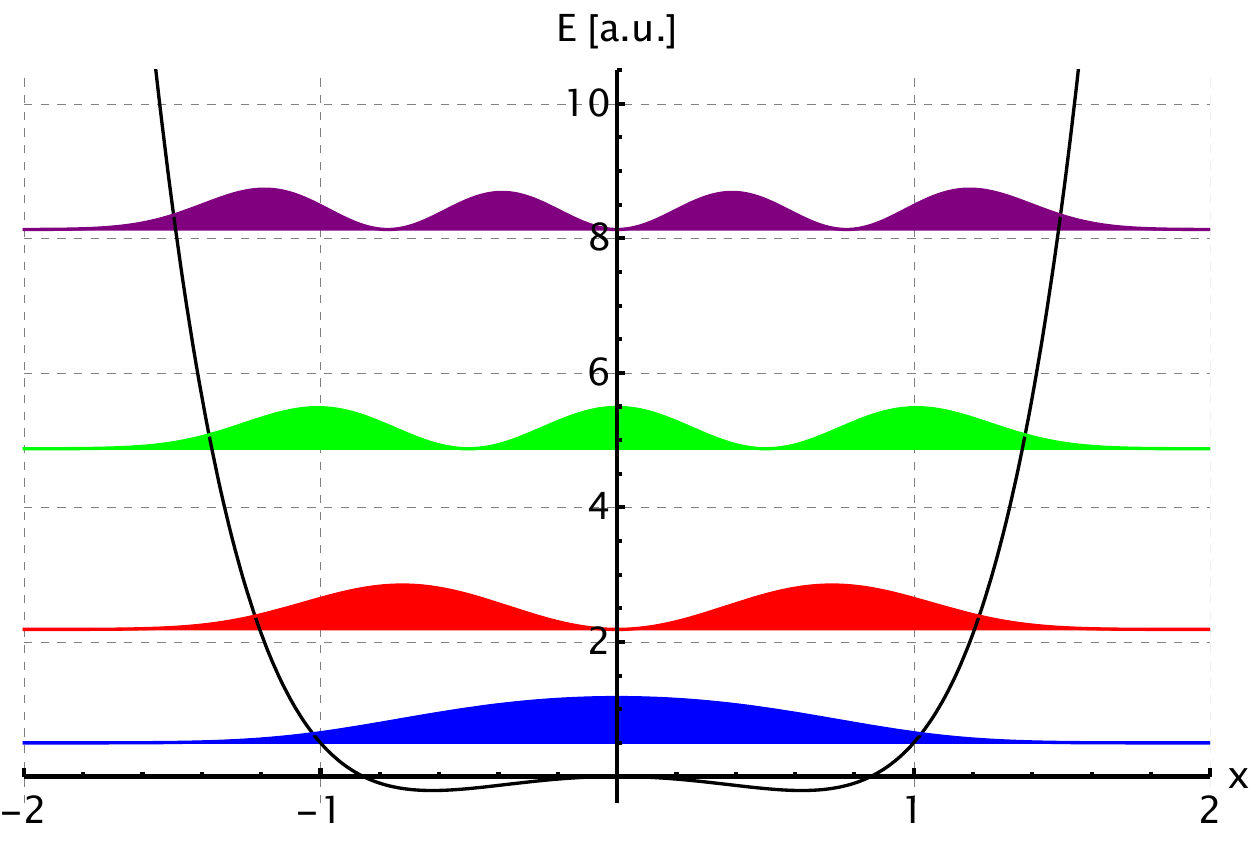}
        \caption*{(a) $\lambda=0$}
    \end{minipage}\hfill
    \begin{minipage}{0.45\textwidth}
        \includegraphics[width=\textwidth]{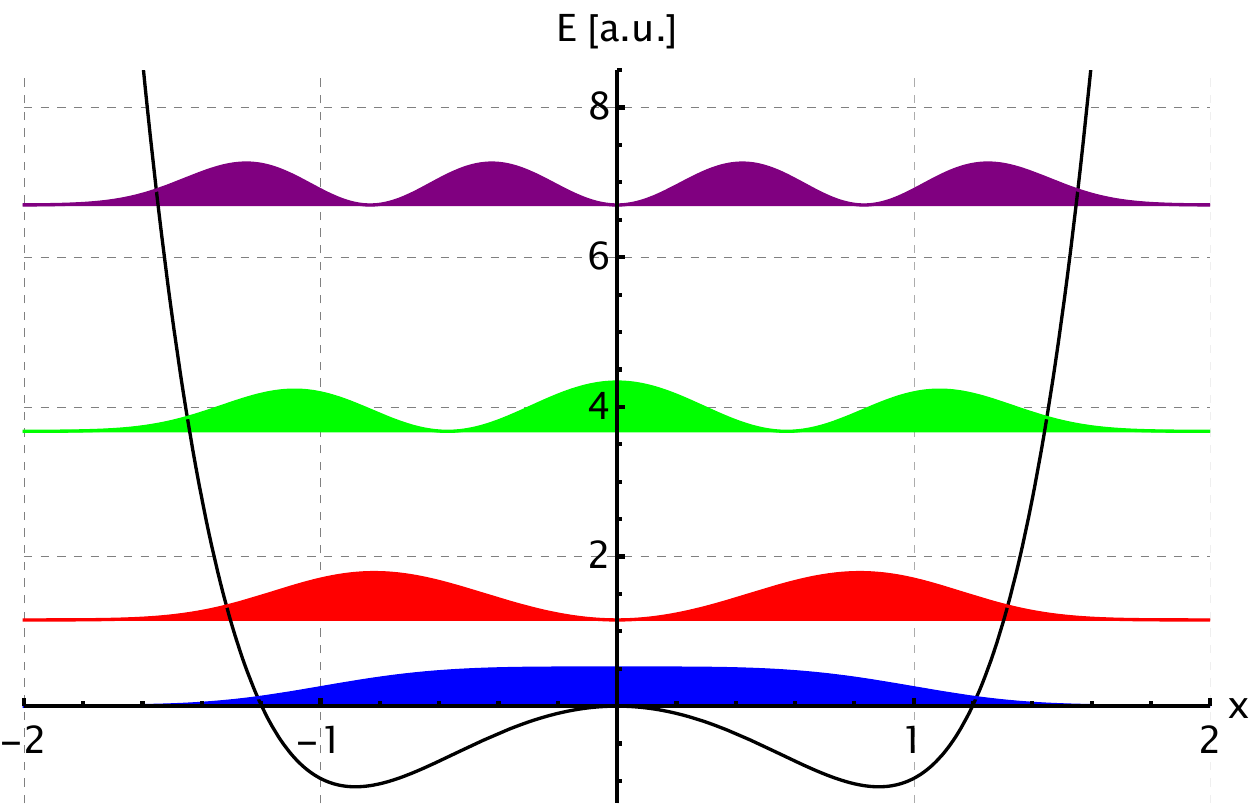}
        \caption*{(b) $\lambda=\lambda_c^{(n=0)}$}
    \end{minipage}\hfill
    \begin{minipage}{0.45\textwidth}
        \includegraphics[width=\textwidth]{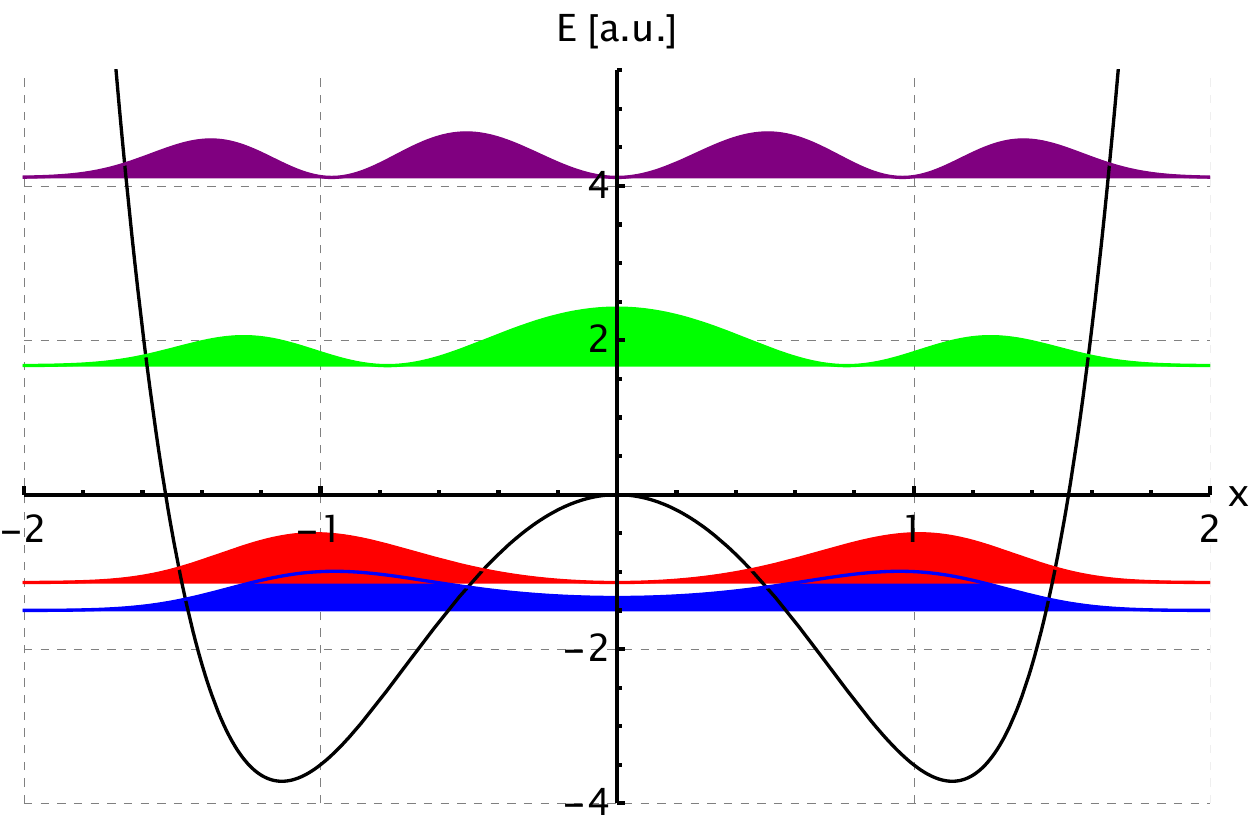}
        \caption*{(c) $\lambda=2$}
    \end{minipage}\\
    \begin{minipage}{0.45\textwidth}
        \includegraphics[width=\textwidth]{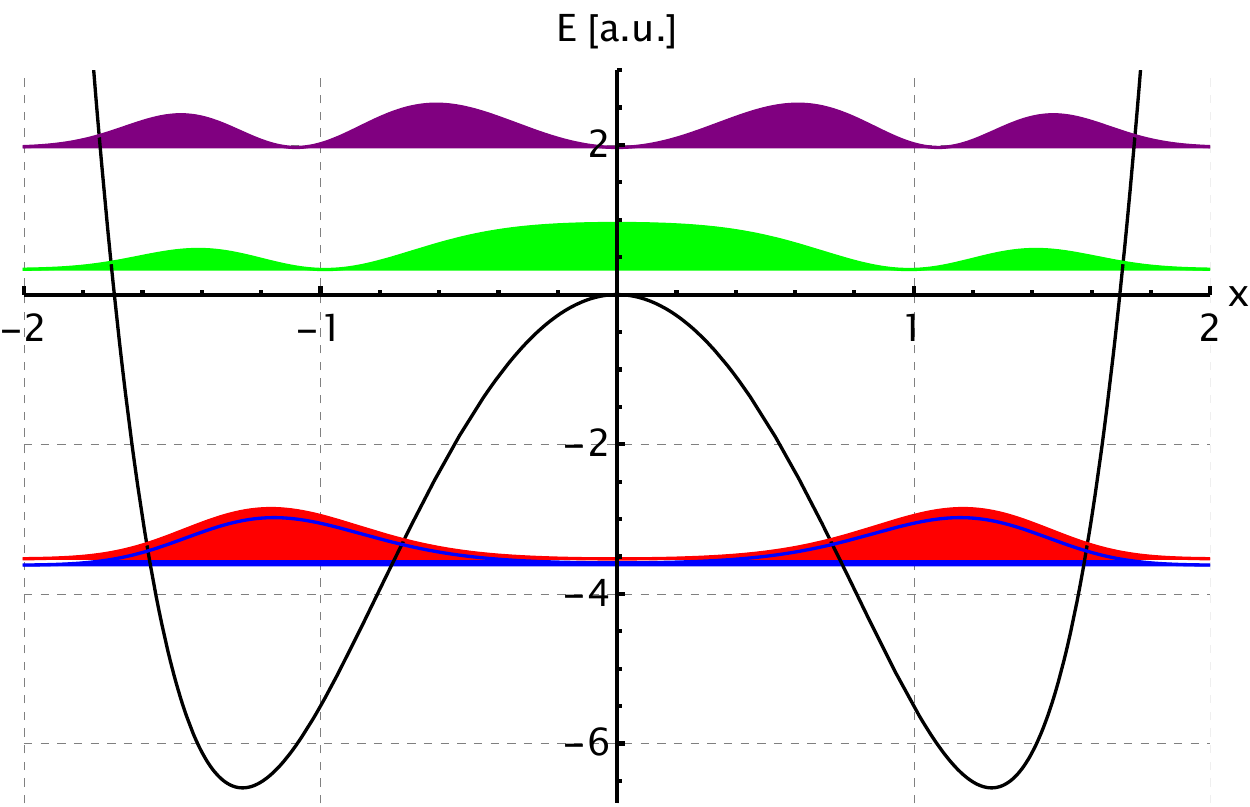}
        \caption*{(d) $\lambda=3$}
    \end{minipage}\hfill
    \begin{minipage}{0.45\textwidth}
        \includegraphics[width=\textwidth]{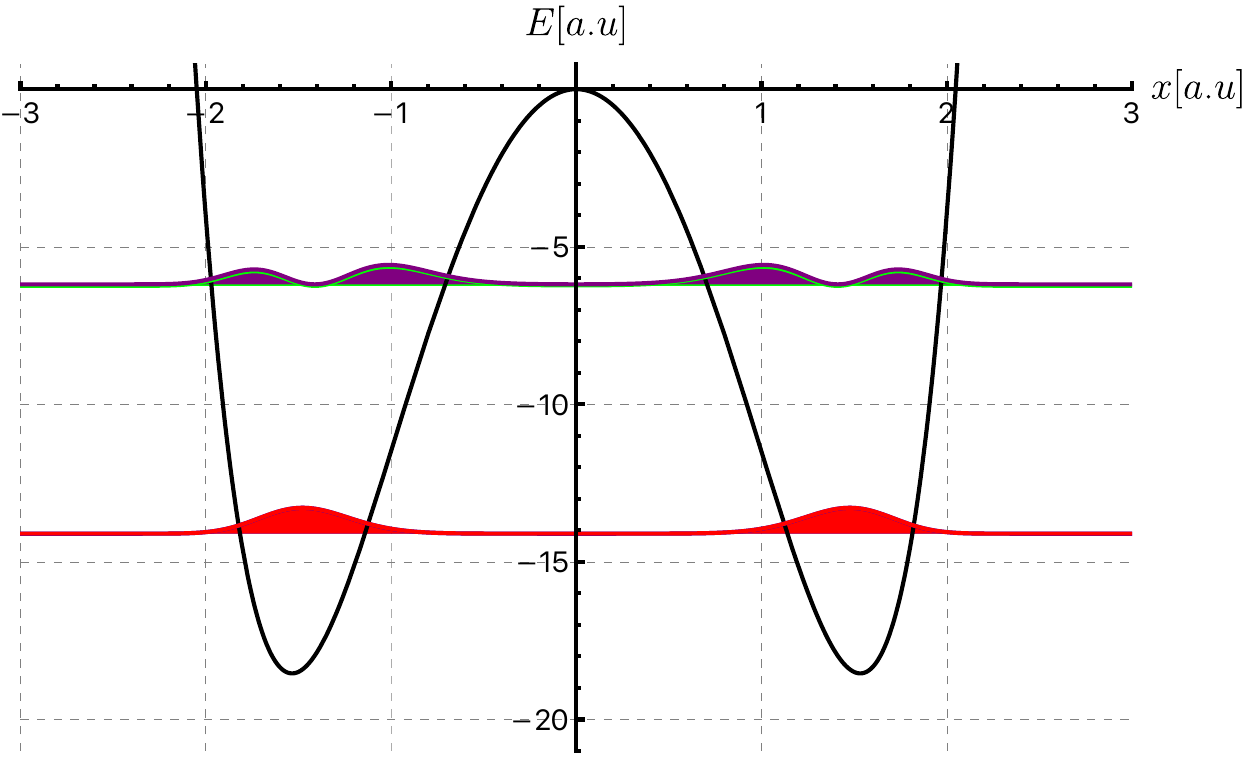}
        \caption*{(e) $\lambda=6$}
    \end{minipage}
    \caption{\small The QES potential \( V^{\rm QES} \) (\ref{vqes}) (black curve) and the lowest energy levels \( n = 0, 1, 2, 3 \) at various \( \lambda \). As \( \lambda \) increases, the potential transitions from a single to a double well, producing level splitting, barrier penetration, and eventual symmetry breaking with localized doublets.
}
    \label{Vpsi}
\end{figure}

Figure~\ref{Vpsi} exhibits the evolution of the QES sextic potential \( V_{\text{QES}}(x) \), shown as a black curve, along with the lowest four energy levels and corresponding {probability densities} for various values of the parameter \( \lambda \). Each panel displays the potential together with the squared moduli of the wavefunctions \( |\psi_n(x)|^2 \) (colored bands), vertically aligned with their respective energy eigenvalues for \( n = 0, 1, 2, 3 \).

For negative and small positive values of \( \lambda \) (e.g., \( \lambda = -3/4 \) and { \( \lambda = 0 \)}), the potential exhibits effectively a single-well structure and the eigenfunctions are delocalized, with all energy levels lying above the classical barrier at \( E = 0 \). No exact solutions are known for $\lambda< 0$. As \( \lambda \) increases, the potential develops a pronounced double-well structure. At the critical value \( \lambda = \lambda_c^{(n=0)} \), the ground state energy crosses the barrier height, marking the onset of tunneling. 

For \( \lambda = 0 \), the potential admits one quasi-exact solution, the ground state which lies above the barrier. At \( \lambda = 2 \) and \( \lambda = 3 \), three and four exact eigenstates exist. In these cases, the pair of wavefunctions $n=0,1$ are well-localized in each well, and the emergence of near-degenerate level pairs becomes evident.

By $\lambda = 6$, the double-well structure is fully developed. The potential supports multiple bound states below the central barrier, and quasi-degenerate level pairs are clearly visible in the low-energy part of the spectrum.

This sequence of panels demonstrates how tuning \( \lambda \) not only affects the eigenvalue spectrum but also modifies the potential landscape itself. Thus, the figure represents a one-parameter family of sextic potentials rather than a single fixed potential, emphasizing the connection between the structure of the potential and the state pairing of its eigenstates.

\clearpage

\clearpage

{

\subsubsection{The left- and right-localized states  \( \psi_L(x) \) and \( \psi_R(x) \)}
The two lowest energy eigenstates of a symmetric double well potential are the symmetric and antisymmetric states, denoted by \( \psi_0(x)=\psi_+(x) \) and \( \psi_1(x)=\psi_-(x) \), which are delocalized across both wells. However, when the barrier is sufficiently high (large $\lambda-$limit) and tunneling is weak, these states can be used to construct approximate wavefunctions localized in the left and right wells. The left- and right-localized states \( \psi_L(x) \) and \( \psi_R(x) \) are given by the linear combinations:

\[
\psi_L(x) = \frac{1}{\sqrt{2}} \left( \psi_+(x) + \psi_-(x) \right), \qquad
\psi_R(x) = \frac{1}{\sqrt{2}} \left( \psi_+(x) - \psi_-(x) \right).
\]

These localized states are not eigenfunctions of the Hamiltonian, but they are useful for semiclassical interpretations and for understanding the system's behavior in the limit of negligible tunneling. In this limit, the symmetric and antisymmetric states become nearly degenerate, and the system effectively behaves as if it has two classically isolated wells with distinct localized states.

}

\subsection{Momentum space}

As previously mentioned, the trial functions (\ref{trialf}) we are using possess the property that their {Dirac-Fourier} transform can be obtained analytically. For instance, in the cases $\lambda=0,\frac{1}{2},\lambda_c,\frac{3}{2}$ with $n=0$ (ground state) the optimal variational function in position space is of the form

\begin{equation}
\label{}
    \psi_{0}(x) \ = \ A\,(1\,+\,a_1\,x^2+\,a_2\,x^4+\,a_3\,x^6+\,a_4\,x^8+\,a_5\,x^{10})\,e^{-\frac{x^4}{4}} \ ,
\end{equation}
it contains solely five variational parameters $a_i$ {(Appendix A)}, and $A$ is the normalization constant. The corresponding Fourier transform $\phi(p) = \frac{1}{\sqrt{2\,\pi}} \int_{-\infty}^{\infty} \psi(x) \, e^{-i \,p\, x} \, dx$, the ground state function $\phi_0(p)$ in momentum space, reads
{\small
\begin{equation}
\begin{aligned}
& \phi_0(p) \ =  \  \frac{A\, \Gamma\,\left(\frac{5}{4}\right)}{\sqrt{\pi }}\bigg[\,  6 \left(a_2 \, _1F_3\left(\frac{5}{4};\frac{1}{4},\frac{1}{2},\frac{3}{4};\frac{p^4}{64}\right)+5\, a_4 \, _1F_3\left(\frac{9}{4};\frac{1}{4},\frac{1}{2},\frac{3}{4};\frac{p^4}{64}\right)+45\, a_6 \, _1F_3\left(\frac{13}{4};\frac{1}{4},\frac{1}{2},\frac{3}{4};\frac{p^4}{64}\right)\right)
\\ &
-3 p^2 \left(a_1 \, _0F_2\left(0;\frac{3}{4},\frac{3}{2};\frac{p^4}{64}\right)+5 a_3 \, _1F_3\left(\frac{9}{4};\frac{3}{4},\frac{5}{4},\frac{3}{2};\frac{p^4}{64}\right)+45 a_5 \, _1F_3\left(\frac{13}{4};\frac{3}{4},\frac{5}{4},\frac{3}{2};\frac{p^4}{64}\right)\right)
\\ &
+6 \, _0F_2\left(0;\frac{1}{2},\frac{3}{4};\frac{p^4}{64}\right) \,\bigg]  +  \frac{A\, \Gamma\,\left(\frac{7}{4}\right)}{3\,\sqrt{\pi }}\bigg[  -2 p^2 \left(a_6 \left(p^4+195\right)-a_3 p^2-14 a_5 p^2+3 a_2+19 a_4+1\right) \, _0F_2\left(0;\frac{5}{4},\frac{3}{2};\frac{p^4}{64}\right)
\\ &
+\ 4\, \left[\,-\left(a_4+18 a_6\right) p^2+a_1+3 a_3+21 a_5\,\right] \, _0F_2\left(0;\frac{1}{4},\frac{1}{2};\frac{p^4}{64}\right)
\\ &
- \ \frac{1}{15}\, p^6 \,\left(-a_5 p^2+a_2+7 a_4+75 a_6\right) \, _0F_2\left(0;\frac{9}{4},\frac{5}{2};\frac{p^4}{64}\right)\, \bigg]
\ ,
\end{aligned}  
\end{equation}
}
where the generalized hypergeometric function ${}_pF_q$ is defined by the power series
\[
{}_pF_q\left(c_1, \ldots, c_p; r_1, \ldots, r_q; z\right) = \sum_{j=0}^{\infty} \frac{(c_1)_j (c_2)_j \cdots (c_p)_j}{(r_1)_j (r_2)_j \cdots (r_q)_j} \frac{z^j}{j!},
\]
here \((c)_j\) denotes the Pochhammer symbol (rising factorial) defined by
\[
(c)_j = c (c+1) (c+2) \cdots (c+j-1), \quad (c)_0 = 1 \ ,
\]
and $\Gamma(s)$ stands for the Gamma function.

The wavefunction \(\phi_0(p)\) remains finite at \(p = 0\), as indicated by the analytic structure of the generalized hypergeometric functions and the absence of singularities in the series expansion near the origin. This regular behavior is typical of ground states in symmetric potentials, where the momentum-space wavefunction usually attains a maximum at \(p = 0\).

Although the exact asymptotic behavior at large momentum is nontrivial, the presence of high-order terms such as \(p^6\), along with the exponential-like suppression arising from the arguments \(p^4/64\) of the hypergeometric functions, suggests that \(\phi_0(p)\) decays rapidly as \(|p|\) increases. This is consistent with the expected localization of bound states in position space for confining potentials. {To our knowledge, this is among the few examples where physically meaningful variational trial functions admit exact analytical Fourier transforms, enabling more reliable entropy computation.}

The existence of an exact analytic expression for \(\phi_0(p)\) enables precise evaluations of quantum information quantities, including momentum-space Shannon entropy and uncertainty products. In addition, it provides a valuable benchmark for testing the accuracy of variational methods and semiclassical approximations within the framework of quasi-exactly solvable systems. {These remarks are applicable to excited states as well. Moreover, for the (variational) first excited state function
\begin{equation}
\label{}
    \psi_{1}(x) \ = \ {\cal A}\,(x\,+\,b_1\,x^3+\,b_2\,x^5+\,b_3\,x^7+\,b_4\,x^9+\,b_5\,x^{11}+\,b_6\,x^{13})\,e^{-\frac{x^4}{4}} \ ,
\end{equation}
the Dirac- Fourier transform
$\phi_1(p)$ in momentum space is purely imaginary. Explicitly,
\begin{equation}
\begin{aligned}
& \phi_1(p) \ =  \  -\frac{i\, p\,{\cal A}}{504 \,\sqrt{\pi }}\bigg[\, -42 p^2 \left(b_6 \left(p^4+453\right)-b_3 p^2-18 b_5 p^2+5 b_2+39 b_4+1\right) \, _0F_2\left(0;\frac{3}{2},\frac{7}{4};\frac{p^4}{64}\right)
\\ &
+ \ 252 \left(-\left(\left(b_4+22 b_6\right) p^2\right)+b_1+5 b_3+45 b_5\right) \, _0F_2\left(0;\frac{1}{2},\frac{3}{4};\frac{p^4}{64}\right)
\\ &
+ \ p^6 \left(b_5 p^2-b_2-9 b_4-111 b_6\right) \, _0F_2\left(0;\frac{5}{2},\frac{11}{4};\frac{p^4}{64}\right)  \, \bigg] 
\\ &
+ \ \frac{i\, p\,{\cal A}}{5400\, \sqrt{\pi }}\bigg[\, -540 \left(b_2+14 b_4+221 b_6\right) p^4 \, _0F_2\left(0;\frac{3}{2},\frac{9}{4};\frac{p^4}{64}\right)
\\ &
- \ 5400\, \left(b_6 \left(p^4+231\right)+3 b_2+21 b_4+1\right) \, _0F_2\left(0;\frac{1}{2},\frac{5}{4};\frac{p^4}{64}\right) \ + \ b_5\, p^{10} \, _0F_2\left(0;\frac{13}{4},\frac{7}{2};\frac{p^4}{64}\right)
\\ &
- \ 10\, \left(b_4+22 b_6\right) p^8 \, _0F_2\left(0;\frac{5}{2},\frac{13}{4};\frac{p^4}{64}\right) \ + \ 90 \,\left(b_3+22 b_5\right) p^6 \, _0F_2\left(0;\frac{9}{4},\frac{5}{2};\frac{p^4}{64}\right)
\\ &
+ \ 2700\, \left(b_1+7 b_3+77 b_5\right) p^2 \, _0F_2\left(0;\frac{5}{4},\frac{3}{2};\frac{p^4}{64}\right) \, \bigg] \ .
\end{aligned}  
\end{equation}
In general, for the QES system with even \( n \), the Dirac-Fourier transform is real, while for odd \( n \), it is purely imaginary; a behavior analogous to that observed in the simple harmonic oscillator.

}

\begin{figure}[h]
\subfloat[Ground state $n=0$]{\includegraphics[scale = 0.42]{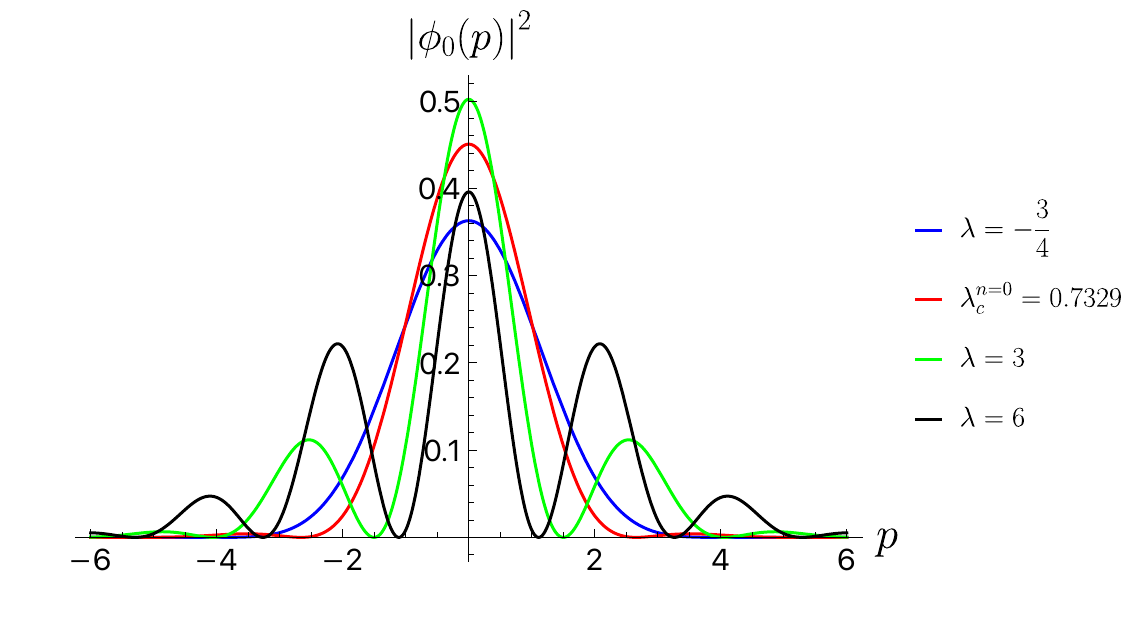}}
\hspace{0.2cm}
\subfloat[First excited state $n=1$]{\includegraphics[scale = 0.42]{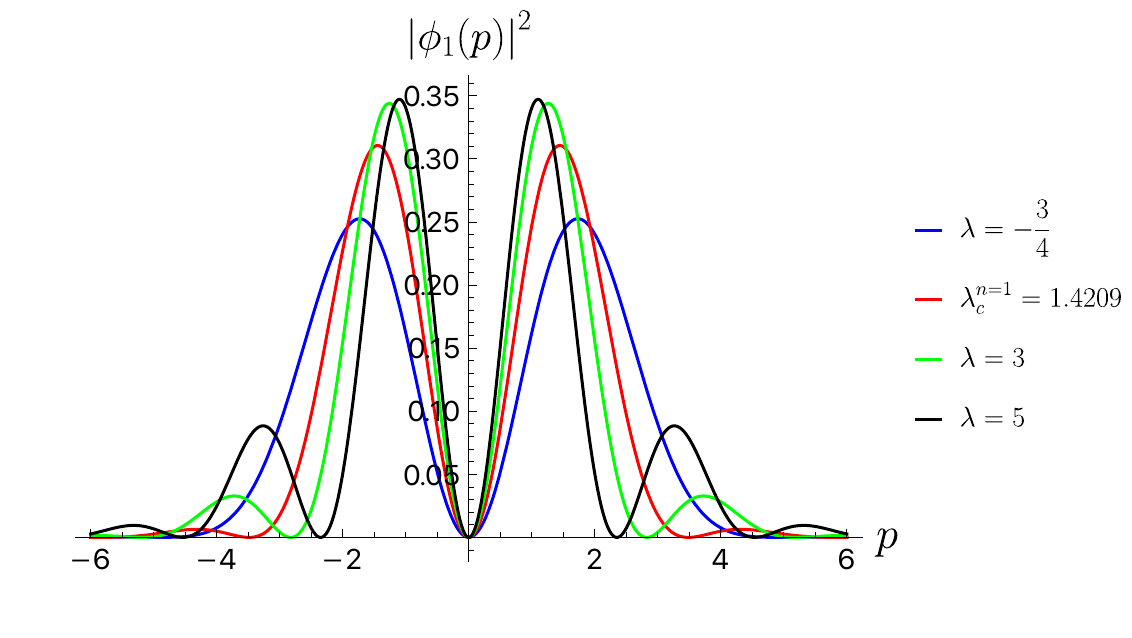}}
\hspace{0.2cm}
\subfloat[Second excited state $n=2$]{\includegraphics[scale = 0.42]{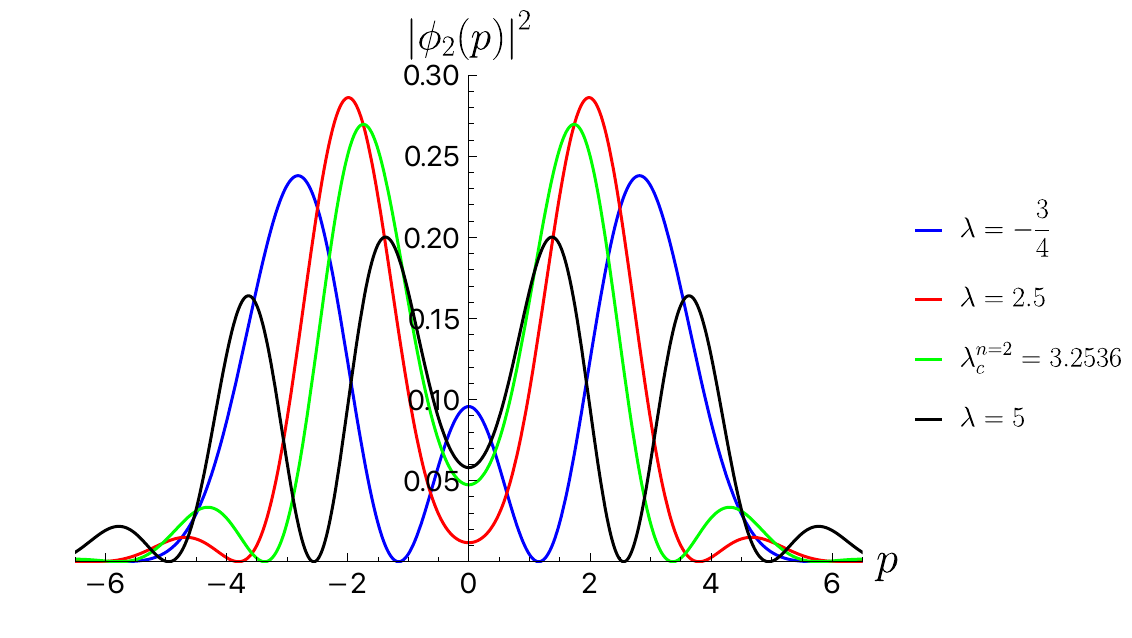}} 
\hspace{0.2cm}
\subfloat[Third excited state $n=3$]{\includegraphics[scale = 0.42]{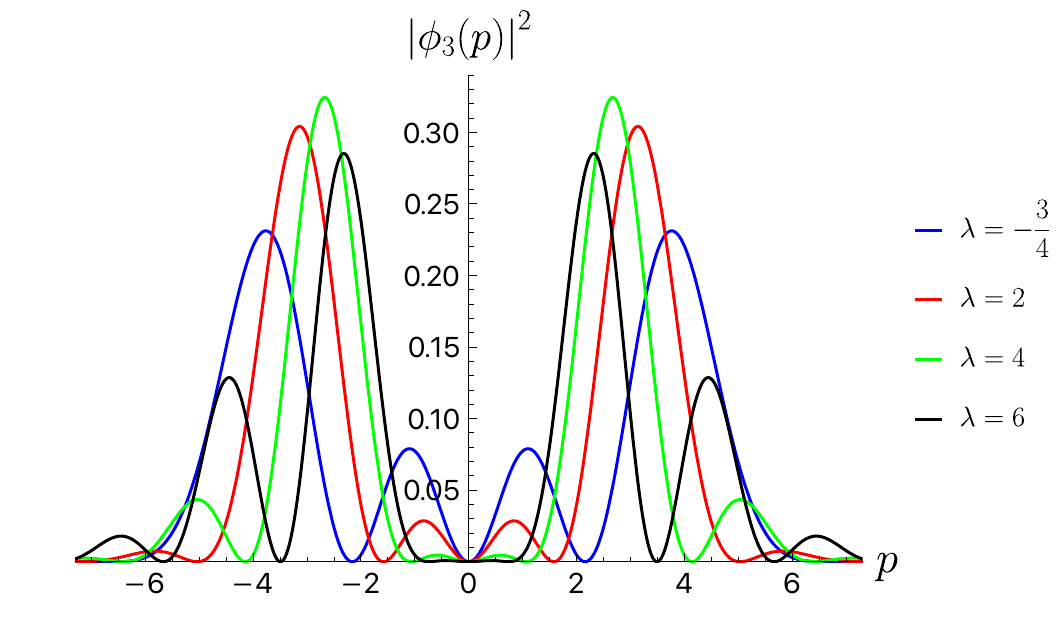}} 
\caption{Momentum-space densities \( |\phi_n(p)|^2 \) for \( n = 0, 1, 2, 3 \) at various \( \lambda \). Even states peak or dip at \( p = 0 \), depending on their nodal structure; odd states vanish there due to antisymmetry. 
}
\label{psimon}
\end{figure}

The Figure~\ref{psimon} displays the momentum-space probability densities $|\phi_n(p)|^2$ for the lowest four eigenstates ($n = 0, 1, 2, 3$) of the QES sextic potential, obtained via Fourier transform of the position-space wavefunctions shown in Figure~\ref{psipos}. These plots illustrate how the structure of the eigenstates in momentum space evolves with the parameter $\lambda$, offering complementary insight into the effects of spatial localization, tunneling, and parity symmetry in the underlying double-well potential.

For $\lambda \geq 1$ the potential deepens into a well-defined double-well structure, the position-space wavefunctions become increasingly localized in the left and right wells as $\lambda$ grows  {(see Figure 5)}. This spatial localization leads to a broadening of the momentum-space profiles due to the uncertainty principle and the emergence of distinct interference patterns. { Panels (a)--(d) reveal oscillatory structures and side-lobes that resemble double-slit interference.} These fringes arise from coherent superpositions of spatially separated lobes and encode the relative phase differences between symmetric and antisymmetric components.

Notably, for states beyond their critical coupling $\lambda_c^{n}$, the momentum-space densities exhibit alternating peaks and dips—\textit{momentum-space nodal splitting}—that grow increasingly complex with higher $n$ and larger $\lambda$. For example, at $\lambda = 5$ or $\lambda = 6$, the second and third excited states ($n = 2, 3$) in panels (c) and (d) show rich fringe patterns with multiple secondary maxima. This behavior reflects strong tunneling coherence and quasi-degenerate level pairing, in which symmetric and antisymmetric combinations of spatially localized states give rise to structured momentum distributions.

\clearpage

Even for nearly degenerate low-lying states, such as $\psi_0(x)$ and $\psi_1(x)$ at $\lambda = 3$ or $\lambda = 5$, the momentum-space densities remain distinguishable. The ground state retains a central peak, while the first excited state shows a central dip due to its odd parity—mirroring their position-space symmetry and confirming that momentum-space observables remain sensitive to parity and coherence even in the tunneling-dominated regime.

Finally, the Figure~\ref{psipgen} presents the momentum-space probability densities $|\phi_n(p)|^2$ for the eigenstates pairs $n = 0, 1$ and $n=2, 3$ at fixed values of $\lambda = 0, 2.5, 3.5, 6$. Each row corresponds to a different value of $\lambda$, while each column groups states by adjacent energy levels: $n = 0, 1$ on the left and $n = 2, 3$ on the right. These momentum-space profiles are obtained via Fourier transform of the position-space wavefunctions and offer insight into the symmetry, localization, and interference effects associated with quantum tunneling.

At $\lambda = 0$, where the potential resembles a single-well, the wavefunctions are relatively delocalized in position space, and their momentum-space counterparts are broad and smooth. At this stage, the momentum distributions of adjacent states are clearly distinguishable: for example, $\phi_0(p)$ is peaked at the origin, while $\phi_1(p)$ exhibits a node at $p = 0$, consistent with its antisymmetric parity. Similar differences are observed between $\phi_2(p)$ and $\phi_3(p)$, due to their differing nodal structure.

As $\lambda$ increases (middle and bottom rows), and the potential develops into a double-well, the position-space wavefunctions become increasingly localized in the left and right wells. This spatial localization translates, via the uncertainty principle, into a broadening of the momentum-space distributions. Crucially, the momentum-space densities of adjacent eigenstates not only remain distinguishable, but they begin to \textit{diverge more sharply} with increasing $\lambda$. This is evident in the emergence of more pronounced \textit{oscillatory patterns}, \textit{side-lobes}, and \textit{peak shifts} in the momentum profiles. These features are especially visible for $\phi_2(p)$ and $\phi_3(p)$ at $\lambda = 6$, where the differences in their interference structure and peak locations reflect the growing impact of nodal complexity and symmetry breaking.

\begin{figure}[h]
    \centering
    \begin{minipage}{0.45\textwidth}
        \includegraphics[width=\textwidth]{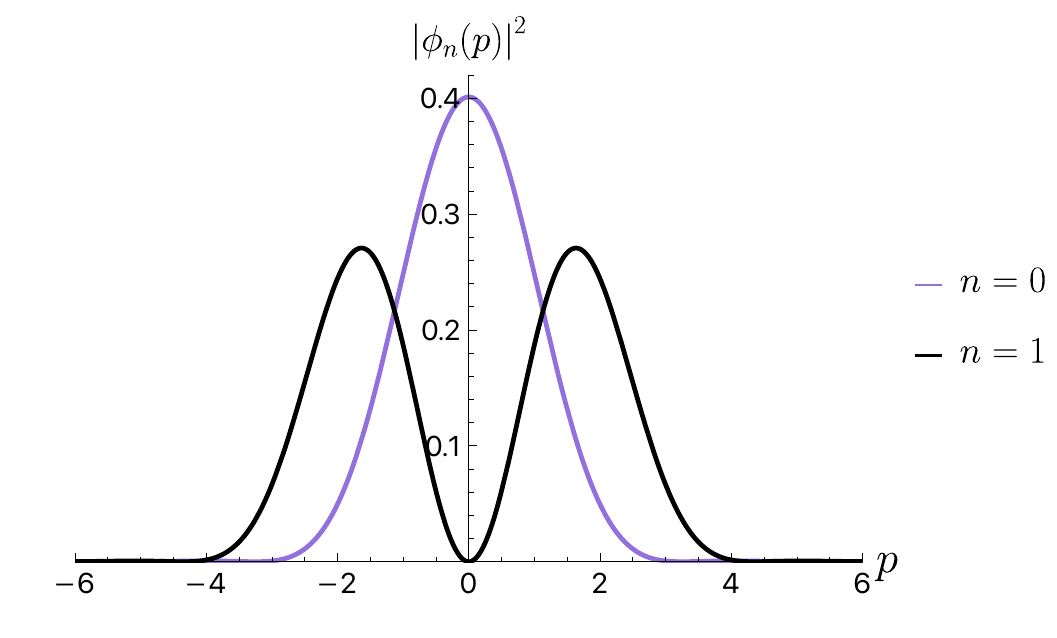}
        \caption*{(a) $\lambda = 0$ fixed}
    \end{minipage}\hfill
    \begin{minipage}{0.45\textwidth}
        \includegraphics[width=\textwidth]{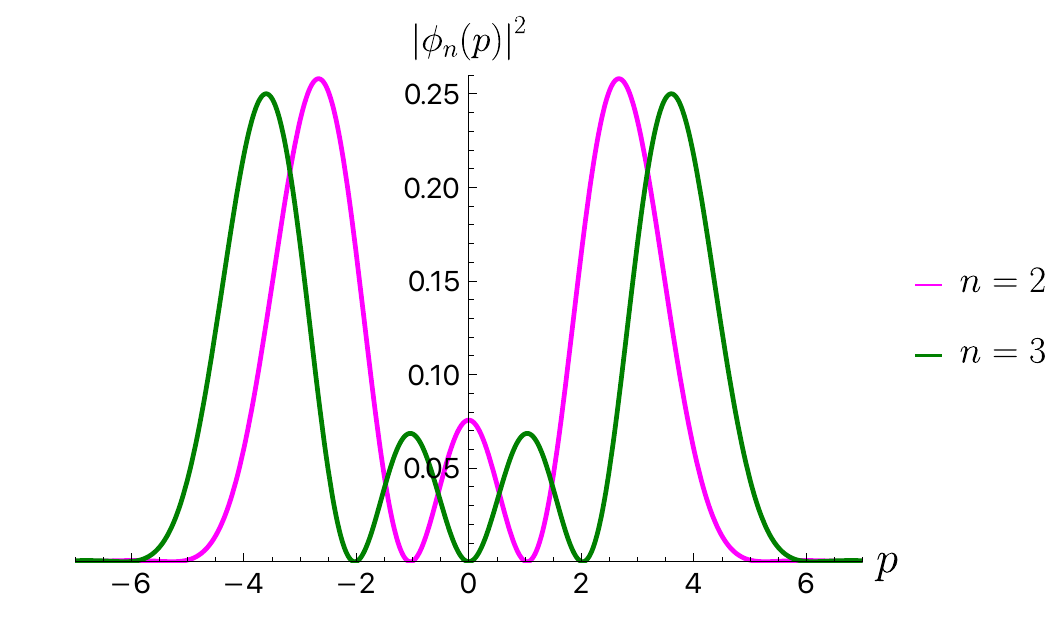}
        \caption*{(b) $\lambda = 0$ fixed}
    \end{minipage}\hfill
    \label{fig:cinco-figuras}
\end{figure}

\begin{figure}[h]
    \centering
    \begin{minipage}{0.45\textwidth}
        \includegraphics[width=\textwidth]{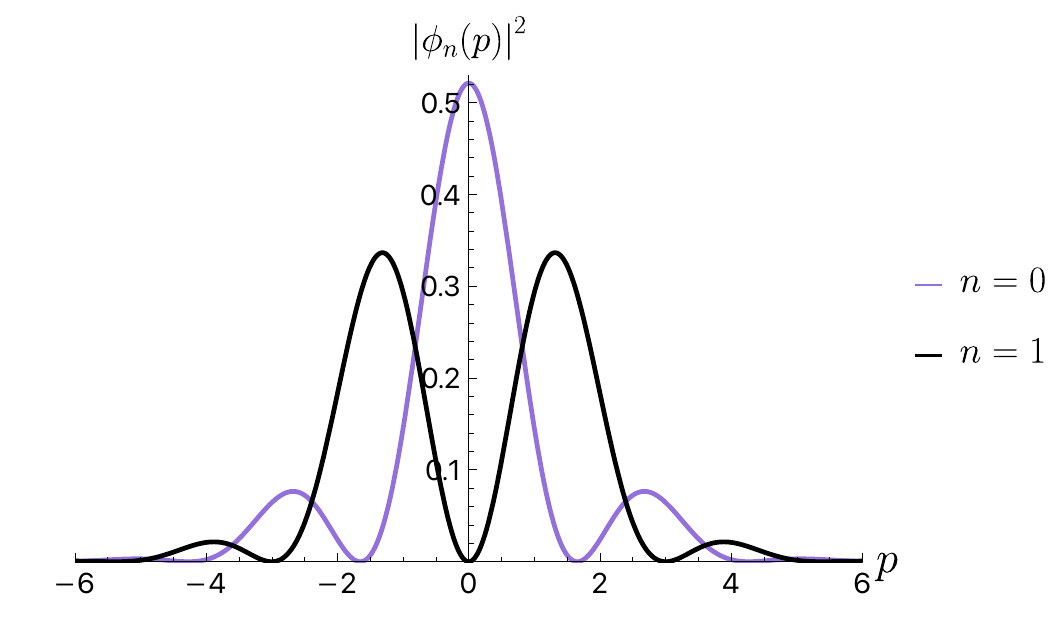}
        \caption*{(c) $\lambda = 2.5$ fixed}
    \end{minipage}\hfill
    \begin{minipage}{0.45\textwidth}
        \includegraphics[width=\textwidth]{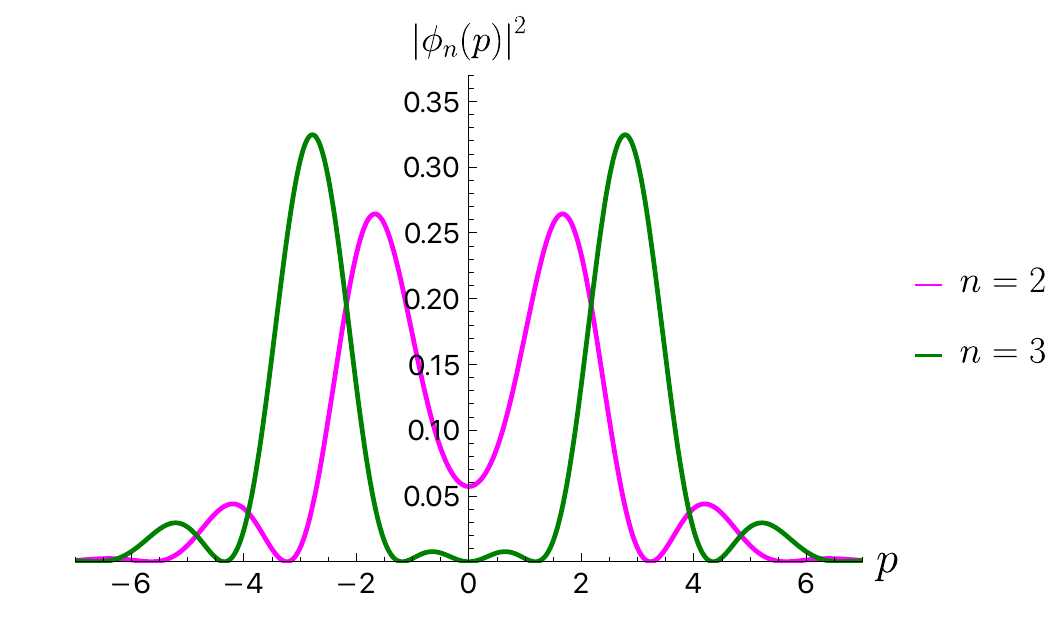}
        \caption*{(d) $\lambda = 3.5$ fixed}
    \end{minipage}\hfill
    \label{fig:cinco-figuras}
\end{figure}

\begin{figure}[h]
    \centering
    \begin{minipage}{0.45\textwidth}
        \includegraphics[width=\textwidth]{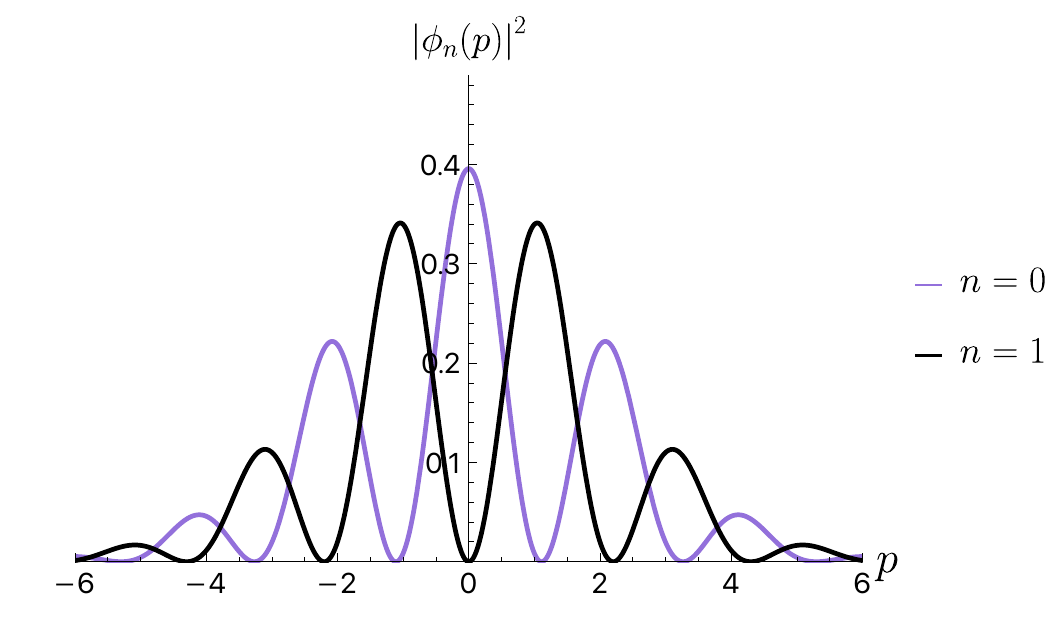}
        \caption*{(e) $\lambda = 6$ fixed}
    \end{minipage}\hfill
    \begin{minipage}{0.45\textwidth}
        \includegraphics[width=\textwidth]{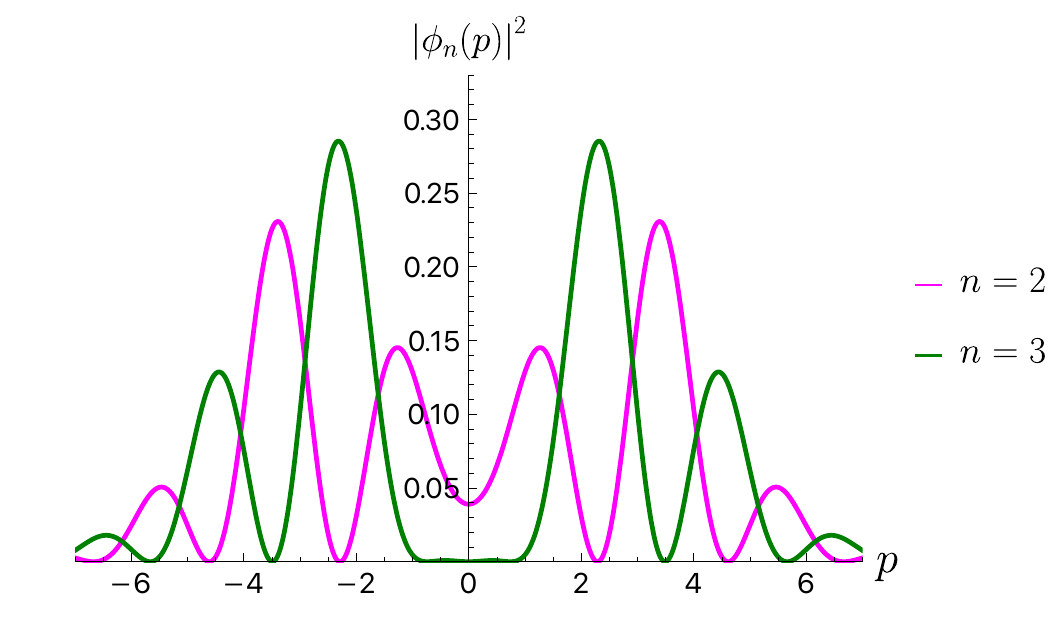}
        \caption*{(f) $\lambda = 6$ fixed}
    \end{minipage}\hfill
\caption{Momentum-space densities \( |\phi_n(p)|^2 \) for \( n = 0,1 \) (left) and \( n = 2,3 \) (right) at \( \lambda = 0, 2.5, 3.5, 6 \). As \( n \) increases, the distributions broaden and become oscillatory due to stronger spatial confinement. Odd states vanish at \( p = 0 \); even excited states show local minima. No quasi-degeneracy appears in momentum space.}
    \label{psipgen}
\end{figure}

\clearpage

This increasing distinction, despite energy levels becoming quasi-degenerate in the tunneling regime, emphasizes the sensitivity of momentum-space observables to the internal structure of the wavefunctions. While in position space, paired states may appear nearly identical aside from a central node, their momentum-space signatures capture subtle differences in phase and coherence due to tunneling-induced symmetry. In particular, antisymmetric states consistently display nodes at $p = 0$, while symmetric ones retain central peaks, a feature that remains robust and sharpens with larger $\lambda$.

Importantly, Figure~\ref{psipgen} shows that \textit{momentum-space densities of adjacent eigenstates remain distinguishable across all coupling regimes and become increasingly different as $\lambda$ increases}. This growing divergence reflects enhanced spatial localization, stronger parity separation, and the emergence of complex interference effects, underscoring the complementary role of momentum-space analysis in understanding tunneling dynamics and wavefunction structure in quasi-exactly solvable systems.

{
\subsubsection{Interference and Pairing in the Momentum-Space Profiles of  \( \psi_n(x) \)}

In a symmetric double well potential, the eigenstates alternate in parity: the wavefunctions \( \psi_n(x) \) for even \( n \) are symmetric under spatial inversion, while those with odd \( n \) are antisymmetric. The ground state \( \psi_0(x) \) and the first excited state \( \psi_1(x) \) form a nearly degenerate tunneling doublet. Their linear combinations yield left- and right-localized states, \( \psi_L(x) = \frac{1}{\sqrt{2}}(\psi_0 + \psi_1) \) and \( \psi_R(x) = \frac{1}{\sqrt{2}}(\psi_0 - \psi_1) \), each predominantly supported in one of the two potential wells. In position space, interference between these components leads to enhanced central density in the symmetric state and a node in the antisymmetric one—hallmarks of constructive and destructive interference, respectively.

This parity structure directly governs the behavior of the Fourier transforms. For even \( n \), the wavefunction \( \psi_n(x) \) is real and even, implying a Fourier transform \( {\phi}_n(p) \) that is \textit{real and even}. For odd \( n \), the wavefunction is real and odd, and its Fourier transform is \textit{purely imaginary and odd}. When combining even and odd states to form localized superpositions, these parity-induced phase differences yield \textit{complex-valued, asymmetric momentum-space wavefunctions}. The resulting interference pattern in momentum space exhibits \textit{twin peaks} centered at \( \pm p_0 \), a manifestation of coherent superposition between counter-propagating momentum components. This phenomenon, often referred to as \textit{momentum-space pairing}, reflects the underlying tunneling-induced coherence between spatially separated wells.

As one proceeds to higher excited states (\( n \geq 2 \)), similar symmetric–antisymmetric doublets persist (e.g., \( \psi_2, \psi_3 \), etc.), and their superpositions continue to exhibit parity-dependent interference patterns. However, these wavefunctions possess more nodes and become increasingly delocalized, with their position-space probability densities showing finer spatial fringes and their Fourier transforms displaying broader structures with weaker pairing. At higher energies, the states begin to access the classically allowed region above the barrier, where tunneling becomes less relevant. In this regime, left- and right-localized interpretations lose physical clarity, and the wavefunctions transition to a \textit{fully delocalized, interference-dominated regime}, characterized by highly oscillatory patterns in both position and momentum space. While parity remains a good quantum number, the clean interference signatures and momentum-space pairing observed in the low-energy doublet gradually vanish as the system approaches semiclassical behavior.

}

\section{Heisenberg and Shannon Entropic uncertainties}
\label{s5}

In quantum mechanics, the uncertainties in position and momentum are defined as
\[
\Delta x = \sqrt{\langle x^2 \rangle - \langle x \rangle^2}, \quad \Delta p = \sqrt{\langle p^2 \rangle - \langle p \rangle^2}.
\]
These quantities measure the standard deviations of the corresponding observables. Their product is constrained by the Heisenberg uncertainty principle,
\[
\Delta x\, \Delta p \ \geq \ \frac{1}{2}\ ,
\]
which establishes a fundamental lower bound on the simultaneous precision with which position and momentum can be determined. States that saturate this inequality, such as Gaussian wave packets (like the ground state of the harmonic oscillator), are referred to as minimum uncertainty states and play a central role in the characterization of quantum coherence and localization phenomena.

In addition to standard deviations, uncertainty in quantum mechanics can also be quantified using \textit{Shannon entropy}. For a normalized wavefunction $\psi(x)$ in position space and its Fourier transform $\phi(p)$ in momentum space, the Shannon entropies are defined as
\[
S_x = -\int |\psi(x)|^2 \ln |\psi(x)|^2\, dx, \quad S_p = -\int |\phi(p)|^2 \ln |\phi(p)|^2\, dp.
\]
These entropies measure the information-theoretic uncertainty associated with the probability distributions in position and momentum space, respectively.

A fundamental \textit{entropic uncertainty relation}, stronger in some respects than the standard deviation  based Heisenberg inequality, places a lower bound on the sum of these entropies: {
\[
S_t \ = \ S_x + S_p \ \geq\ D\,(1+\ln\pi ) \ ,
\]
where $D$ is the spatial dimension (here $D=1$).}

This inequality expresses the fact that high localization in one domain (e.g., position) necessarily implies a spread in the conjugate domain (momentum), in accordance with the wave nature of quantum particles.

As with the standard deviation-based uncertainty relation, \textit{Gaussian wave packets} also saturates this entropic bound. For such states, the probability densities in both position and momentum space are Gaussian, and the sum $S_x + S_p$ reaches the minimum value {$(1+\ln\pi ) \approx 2.1447$}. These states are thus referred to as \textit{minimum entropy states}, reflecting their optimal balance between localization and delocalization.

Entropic uncertainty relations offer a more general framework for quantifying quantum uncertainty, particularly valuable in scenarios where standard deviations are ill-defined or inadequate, such as in the presence of long-tailed or non-Gaussian distributions.

\begin{figure}[h]
\begin{center}
\includegraphics[width=18cm]{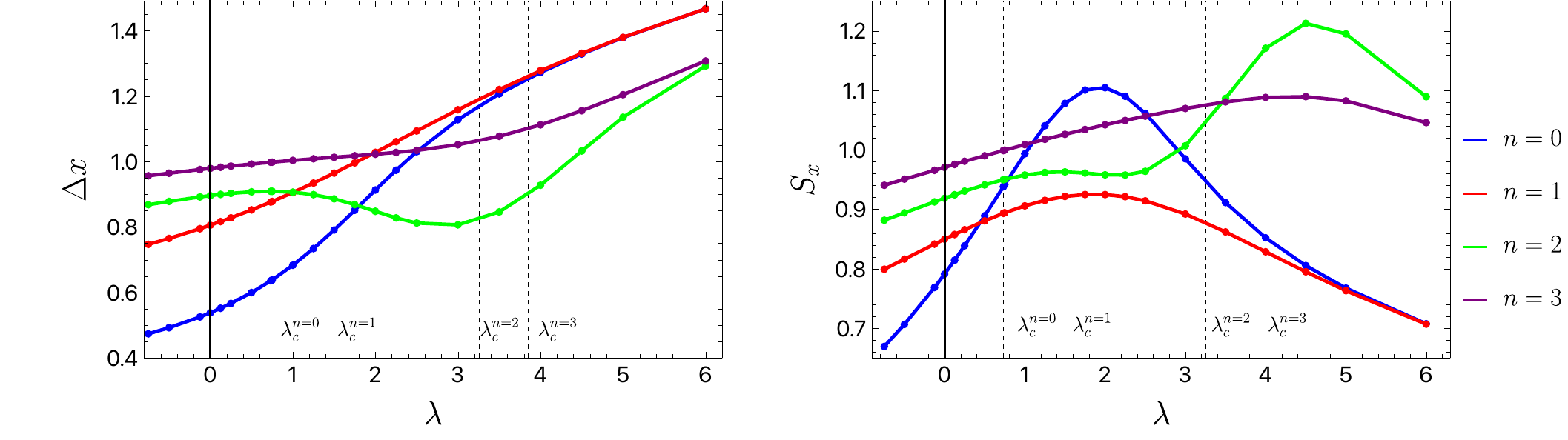}
\caption{\small Position uncertainty \( \Delta x \) (left) and Shannon entropy \( S_x \) (right) vs.\ \( \lambda \) for states \( n = 0,1,2,3 \). While \( \Delta x \) grows smoothly (except for \( n = 2 \)), \( S_x \) shows non-monotonic behavior with peaks near \( \lambda_c^{(n)} \), and for $n=2$ an additional extremum beyond them, indicating structural changes in the wavefunction not captured by variance. Results use the optimal trial function (\ref{trialf}).
}
\label{DeltaxN}
\end{center}
\end{figure}

Figure~\ref{DeltaxN} shows the position uncertainty $\Delta x$ (left panel) and the Shannon entropy in position space $S_x$ (right panel) as functions of the parameter $\lambda$ for the first four eigenstates ($n = 0, 1, 2, 3$) of the sextic QES  potential. For $\lambda > -\tfrac{1}{2}$, the potential develops a double-well structure with a central barrier at $x = 0$. At critical values $\lambda = \lambda_c^n$ (vertical dashed lines), the energy of the $n$-th state crosses the barrier height $E = 0$, marking the transition from single-well localization to double-well delocalization and the onset of tunneling.

The uncertainty $\Delta x$ increases with $\lambda > \lambda_c^n$ for all states, reflecting the increasing depth of the potential wells. In the case $n=2$, $\Delta x$ displays a minimum at $\lambda < \lambda_c^{n=2}$ whereas the states \( n = 0, 1, 3 \) display monotonic profiles. In general, changes in slope near each $\lambda_c^n$ indicate wavefunction spreading across both wells. Notably, curve crossings occur, showing that in certain $\lambda$ regimes, higher excited states may be more localized than lower ones, highlighting tunneling-induced reordering. At large values of \(\lambda\), the pairing between the states becomes evident.

The entropy $S_x$ displays richer structure: each curve exhibits a maximum near after its corresponding $\lambda_c^n$, signaling maximal delocalization and spatial complexity of the wavefunction. These maxima are not seen in $\Delta x$ for $n=0,1,3$, which varies monotonically. For the ground state, the pronounced peak in $S_x$ {after} $\lambda_c^0$ captures the transition to a symmetric superposition spanning both wells, where positional uncertainty is maximized. This contrast underscores the superior sensitivity of entropy-based measures in detecting qualitative changes in wavefunction structure and coherence. Moreover, the value of \( S_x \) reveals that, at \( \lambda = 6 \), the \( n = 2 \) and \( n = 3 \) states possess distinct informational content, despite exhibiting nearly identical spatial widths \( \Delta x \)  {(see Figure \ref{fig:six-panels} (f) and Figure \ref{DeltaxN} (left panel))}. This highlights the presence of structural features beyond mere localization.

Thus, both $\Delta x$ and $S_x$ show the interplay between quantum localization, tunneling, and potential geometry, but $S_x$ provides sharper insight into structural transitions and state reorganization in the QES system.

\begin{figure}[h]
\begin{center}
\includegraphics[width=18cm]{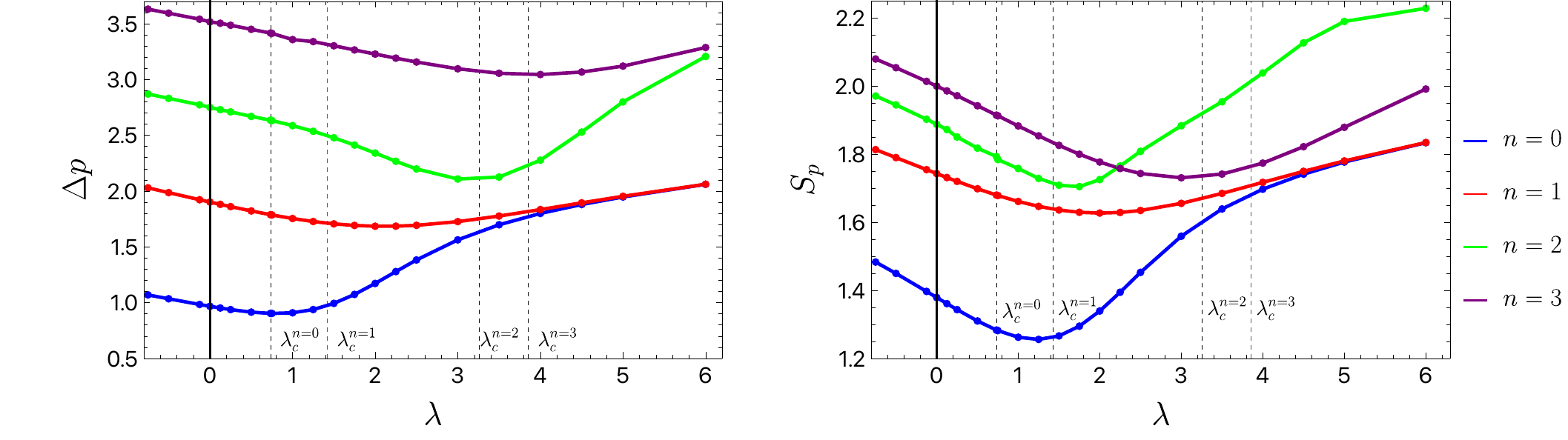}
\caption{Momentum uncertainty \( \Delta p \) (left) and Shannon entropy \( S_p \) (right) vs.\ \( \lambda \) for \( n = 0,1,2,3 \). The rise beyond the critical values \( \lambda_c^{(n)} \), indicate compression and fragmentation in momentum space. They capture similar structural features. In \( \Delta p \) no curve crossings appear, and the ordering with respect to \( n \) is preserved.
}
\label{DeltapN}
\end{center}
\end{figure}

Figure~\ref{DeltapN} presents the momentum uncertainty $\Delta p$ (left panel) and the Shannon entropy in momentum space $S_p$ (right panel) as functions of the parameter $\lambda$, for the first four eigenstates ($n = 0, 1, 2, 3$) of the  QES sextic potential. As $\lambda$ increases and the potential transitions into a double-well configuration for $\lambda > -\tfrac{1}{2}$, both $\Delta p$ and $S_p$ display behavior complementary to their position-space counterparts.

In the left panel, $\Delta p$ decreases with $\lambda$ until reaching a minimum near the critical values $\lambda_c^n$, followed by a gradual increase. This behavior reflects the delocalization of the wavefunction in position space and the corresponding contraction in momentum space, consistent with the uncertainty principle. The minima near $\lambda_c^n$ indicate the point of maximal positional spread due to tunneling, after which the wavefunction becomes more localized within individual wells, leading to an increase in momentum uncertainty.

The entropy $S_p$ in the right panel exhibits non-monotonic behavior, with local minima near each $\lambda_c^n$. These minima mark the points of greatest localization in momentum space, corresponding to the delocalized, coherent superpositions in position space. As with $\Delta p$, the curves for different $n$ may cross, signaling state-dependent restructuring in the momentum representation as the potential changes. Interestingly, no curve crossings occur in $\Delta p$. 

In the large-\( \lambda \) regime, the pairing of states becomes apparent not only in position space but also in their momentum-space widths \( \Delta p \), reflecting the underlying spectral structure.
Again, \( S_p \) shows that, at \( \lambda = 6 \), the \( n = 2 \) and \( n = 3 \) states possess distinct informational content, despite having nearly identical spatial widths \( \Delta p \), indicating structure beyond mere localization.

It is worth noting that the behavior of \( \Delta p \) and \( S_p \) is qualitatively similar, whereas \( \Delta x \) and \( S_x \) do not exhibit such correspondence, indicating that momentum-space measures capture certain structural features more clearly.

\subsection{Heisenberg and entropic inequalities}

\begin{figure}[h]
\begin{center}
\hspace{-0.5cm} \includegraphics[width=18cm]{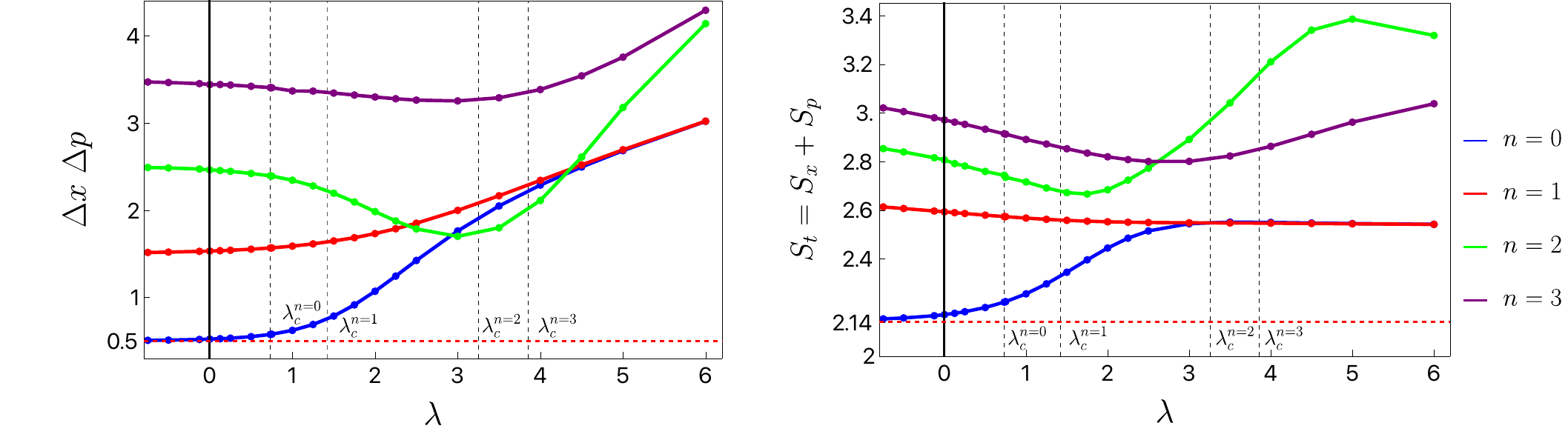}
\caption{\small Heisenberg product 
\( \Delta x \,\Delta p \) (left) and entropic sum 
\( S_t=S_x+S_p \) (right) as functions of $\lambda$ for 
\( n=0,1,2,3 \). Each panel includes its corresponding theoretical lower bound (dashed red line).}
\label{heisenentropic}
\end{center}
\end{figure}

{Figure~\ref{heisenentropic}  illustrates that within the range 
$ -0.75 \leq \lambda \leq 0.2$, the Heisenberg product exhibits behavior characteristic of the harmonic oscillator. In contrast, the entropic sum \( S_t \) deviates from this apparent symmetry, underscoring the greater sensitivity of entropic measures to subtle variations in the underlying potential structure.}

For $\lambda < 0$, the potential behaves effectively as a single-well and is well approximated near the origin by a harmonic oscillator. In this regime, both $\Delta x \Delta p$ and $S_t$ approach their respective lower bounds. In particular, the ground state ($n = 0$) closely saturates the Heisenberg inequality with $\Delta x \Delta p \approx \tfrac{1}{2}$ and reaches the minimal entropy sum $S_t \approx 1+\ln(\pi)$, indicating Gaussian-like, minimum uncertainty behavior.

Interestingly, for the low-lying states $n=0,1,2,3$ the harmonic-like behavior in $\Delta x \Delta p$ persists beyond $\lambda = -\tfrac{1}{2}$, where the potential undergoes a qualitative transition to a double-well structure. Although the potential develops a central barrier and symmetry breaking becomes possible, the states maintain a value of \( \Delta x \, \Delta p \) effectively equal to that of the harmonic oscillator. This is somewhat surprising, as double-well systems typically induce delocalization and bimodal wavefunction shapes. The persistence of low uncertainty values suggests a highly coherent and symmetric ground state that remains delocalized over both wells before significant tunneling asymmetry sets in. We will return to this point later.

As $\lambda$ increases further, the onset of tunneling becomes more pronounced, particularly beyond the critical points $\lambda = \lambda_c^n$ where the energy of the $n$-th state crosses the central barrier height $E = 0$. In this regime, both $\Delta x \Delta p$ and $S_t$ increase significantly, reflecting enhanced delocalization and complexity in the wavefunction.

Crossings among the uncertainty curves are also observed, especially for higher excited states. These crossings indicate a reordering of spatial and momentum uncertainty between different eigenstates as the system evolves. For example, a lower-lying state (e.g., $n=1$) may temporarily become more localized than a higher state ($n=2$) due to tunneling-induced redistribution. 

Moreover, the curves show signs of \textit{pairing behavior}, particularly in the Heisenberg uncertainty panel, where states with adjacent quantum numbers ($n=0$ and $n=1$, $n=2$ and $n=3$) exhibit closely tracking curves beyond their respective critical values. This reflects the formation of nearly degenerate symmetric and antisymmetric combinations of states localized in opposite wells, a hallmark of the tunneling regime. These state pairings correspond to coherent superpositions across the double-well and are tied to the splitting of energy levels due to barrier penetration.

Together, the behavior of $\Delta x \Delta p$ and $S_t$ encapsulates the rich structural transitions of the QES potential, capturing the crossover from harmonic-like confinement to delocalized, symmetry-sensitive tunneling dynamics. The entropy-based measure in particular highlights subtle changes in coherence and localization that are less apparent in the Heisenberg uncertainty product alone.

\section{Quantifying Tunneling and Pairing in the QES Potential Using KL and CRJ Divergences}
\label{s6}

In the study of tunneling dynamics in one-dimensional quantum systems, such as those governed by quasi-exactly solvable sextic double-well potentials, information-theoretic divergences provide valuable tools for quantifying differences between quantum states. Two particularly insightful measures in this context are the Kullback–Leibler (KL) divergence and the Cumulative Residual Jeffreys (CRJ) divergence, both applied to position-space probability densities. Given two normalized densities \( \rho_n(x) \) and \( \rho_m(x) \), corresponding to distinct eigenstates of the system, the KL divergence {from \( \rho_n \) to \( \rho_m \) is defined by
\[
D_{\mathrm{KL}}(\rho_m \parallel \rho_n) = \int_{-\infty}^{\infty} \rho_m(x) \ln\left( \frac{\rho_m(x)}{\rho_n(x)} \right) dx,
\]
which quantifies the information loss when \( \rho_m \) is used to approximate \( \rho_n \). It is asymmetric and becomes infinite if \( \rho_n(x) = 0 \) where \( \rho_m(x) \neq 0 \), making it highly sensitive to mismatches in the support and nodal structure of the distributions.
}

The CRJ divergence instead compares the cumulative residual (or survival) distributions \( S_n(x) = \int_x^{\infty} \rho_n(t) dt \) and \( S_m(x) = \int_x^{\infty} \rho_m(t) dt \), yielding the symmetric measure
\[
D_{\mathrm{CRJ}}(\rho_n \parallel \rho_m) = \int_{-\infty}^{\infty} S_n(x) \ln\left( \frac{S_n(x)}{S_m(x)} \right) dx + \int_{-\infty}^{\infty} S_m(x) \ln\left( \frac{S_m(x)}{S_n(x)} \right) dx.
\]
This formulation ensures that \( D_{\mathrm{CRJ}} \) remains finite even when the densities exhibit non-overlapping support or distinct nodal arrangements, since it smooths out localized zeros via integration.

The structure and distribution of nodes (i.e. points where \( \rho_n(x) = 0 \)) strongly influence both divergence measures, but in distinct ways. The KL divergence is acutely affected by nodal mismatches: if \( \rho_m(x) \) is nonzero where \( \rho_n(x) \) vanishes, the integrand diverges, often rendering \( D_{\mathrm{KL}} \) ill-defined. This limits its applicability when comparing states with differing numbers or placements of nodes, as commonly occurs in quasi-exactly solvable systems. In contrast, the CRJ divergence is more robust to nodal structure. Nodes in the density translate into inflections in the cumulative survival function rather than discontinuities, allowing \( D_{\mathrm{CRJ}} \) to remain well-behaved and sensitive to differences in the spatial extent of tunneling tails.

Thus, while both divergences provide insight into the localization properties of eigenstates, their differing sensitivities make them complementary \cite{HSR2019} (see also \cite{laguna2022information, Hôqua.560560811}). The KL divergence emphasizes disparities in regions of high probability density, typically near classical turning points, while the CRJ divergence accentuates differences in the decay and spread into classically forbidden regions, directly reflecting tunneling behavior. Together, these measures offer a nuanced view of quantum state structure in complex potential landscapes.

\begin{figure}[h]
\begin{center}
\includegraphics[width=11.0cm]{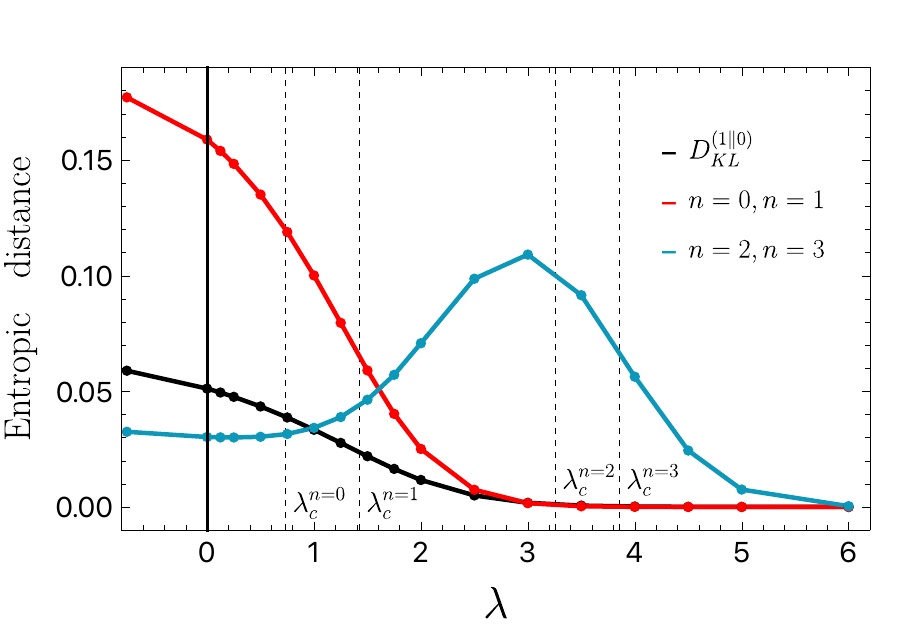}
\caption{ {{CRJ Divergence}  for the eigenstate pairs $(n=0,1)$ {(red curve)}, $(n=2,3)$ {(blue curve)} {and KL Divergence for the eigenstate pairs $(n=0,1)$ (black curve)}versus $\lambda$.}}
\label{CRJ}
\end{center}
\end{figure}

Figure~\ref{CRJ} shows the Cumulative Residual Jeffreys (CRJ) divergence \cite{HSR2019} as a function of the parameter $\lambda$, comparing the spatial probability densities of the eigenstate pairs $(n=0,1)$ ({red curve}) and $(n=2,3)$ ({blue curve}). 

The red curve, corresponding to the pair $(n=0,1)$, starts at a relatively high value for $\lambda < 0$, reflecting the strong distinction between the ground and first excited states in the single-well regime. As $\lambda$ increases and the potential develops a double-well structure, the CRJ divergence decreases sharply and reaches near-zero values for $\lambda \gtrsim 2$. This behavior reflects the emergence of state pairing due to quantum tunneling: the ground and first excited states form symmetric and antisymmetric combinations of nearly identical spatial distributions, leading to near-indistinguishability in their probability densities. The vanishing of the CRJ divergence quantitatively confirms this pairing transition, a hallmark of coherence in double-well systems.

{Notably, the Kullback–Leibler (KL) divergence, represented by the black curve, exhibits a monotonic decrease as a function of $\lambda$, approaching zero for $\lambda \gtrsim 3$. This monotonic trend signals a growing similarity between the ground and first excited states. In the deep double-well regime, this observation is consistent with the CRJ divergence results (see Fig. 11), where both states form a nearly degenerate pair consisting of symmetric and antisymmetric combinations of localized wavefunctions in each well. Since the probability densities—being quadratic in the wavefunction—are nearly identical in this regime, the KL divergence between them also vanishes. This behavior confirms the formation of tunneling-induced state pairs, a hallmark of quantum systems governed by symmetric double-well potentials.}

In contrast, the blue curve shows the CRJ divergence for the higher pair $(n=2,3)$, which begins at low values and remains relatively flat in the single-well regime. Around $\lambda \approx { \lambda_{c}^{n=0}}$, the divergence begins to increase, reaching a maximum near $\lambda \approx 3.2$, and then decreases again. The appearance of this maximum corresponds to the onset of pairing between $n=2$ and $n=3$: prior to this point, the states are structurally distinct, but as the double-well deepens, they begin to form another near-degenerate symmetric-antisymmetric pair. The peak in the CRJ divergence indicates the point of maximal asymmetry in their spatial structures, typically when one state begins to delocalize across both wells while the other remains nodal or partially localized. After the maximum, the decline in CRJ signals increasing similarity as the pairing becomes more complete {(see in Figure \ref{fig:six-panels} (d) and (f)\,)}.

Summarizing, the CRJ divergence serves as a highly sensitive indicator of the tunneling-induced pairing phenomenon in quasi-exactly solvable systems. The suppression of CRJ for $(n=0,1)$ and the non-monotonic behavior for $(n=2,3)$ highlight the critical role of potential geometry and symmetry in reorganizing quantum states. The presence of well-defined minima (or plateaus) at small values confirms the formation of coherent, nearly degenerate doublets, while the maxima reveal regions of rapid structural differentiation—often coinciding with the crossing of critical values $\lambda_c^n$ and associated wavefunction delocalization.

\begin{figure}[h]
\begin{center}
\includegraphics[width=11.0cm]{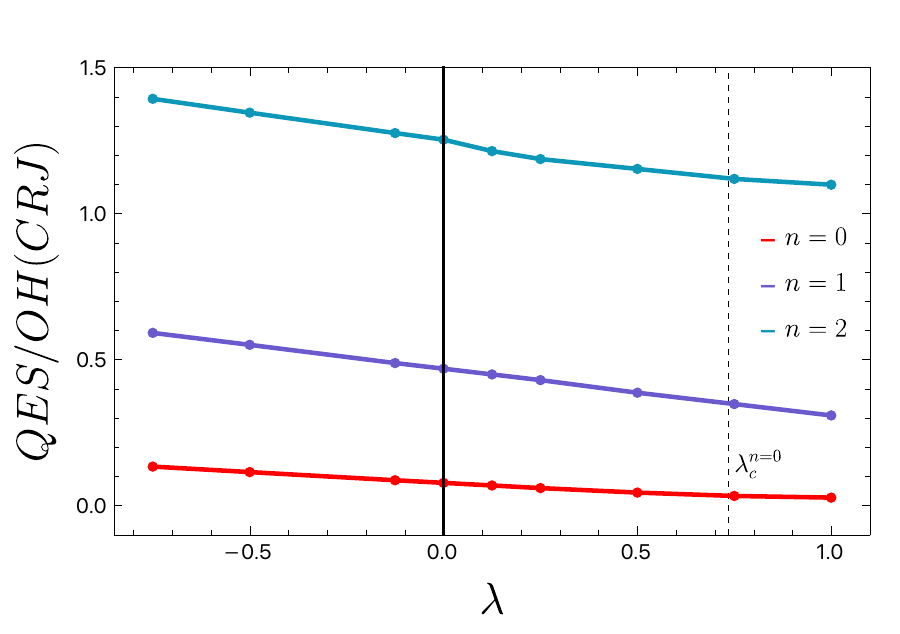}
\caption{\small CRJ divergence between QES sextic and harmonic oscillator densities for \( n = 0, 1, 2 \) \textit{vs} \( \lambda \). Despite similar Heisenberg uncertainty products, the QES states exhibit nontrivial structural deviations from their HO counterparts.}
\label{fig:crj_ratio_qes_ho}
\end{center}
\end{figure}

To assess the degree of similarity between the quasi-exactly solvable (QES) sextic eigenstates and their harmonic oscillator (HO) counterparts, we compute the Cumulative Residual Jeffreys (CRJ) divergence between the position-space probability densities \( \rho_n^{\mathrm{QES}}(x) \) and \( \rho_n^{\mathrm{HO}}(x) \). Figure~\ref{fig:crj_ratio_qes_ho} displays the corresponding CRJ divergence ratios as functions of the parameter \( \lambda \) for the lowest three quantum states (\( n = 0, 1, 2 \)).

While the Heisenberg uncertainty product remains approximately constant (matching the value of the harmonic oscillator) across this interval of \( \lambda \), the CRJ measure reveals significant deviations in the detailed structure of the corresponding densities, particularly in the tails. These differences become more pronounced with increasing \( n \), reflecting enhanced tunneling or asymmetry in the QES states relative to the HO basis. 

These contrasting behaviors underscore the distinct ways in which QES states reorganize in response to the double-well transition. In general, for $\lambda>1$, while some states regain harmonic-like features due to symmetric delocalization, others increasingly depart due to node formation, asymmetry, or partial localization. A brief summary of the main findings of the present study is presented in Table \ref{Tsumm}. Finally, it is worth mentioning that the CRJ divergence (of the order $\sim 10^{-10}$) between the available exact QES probability densities and their variational approximation confirms the accuracy and reliability of our trial function (\ref{trialf}), see Table \ref{TCRJQes}.

\begin{table}[ht]
\centering
\begin{tabular}{|c|c|c|}
\hline
\hspace{0.2cm} $\lambda$ \hspace{0.2cm}  &  \hspace{0.2cm}  state  \hspace{0.2cm}  &  \hspace{0.2cm} $D_{\mathrm{CRJ}}(\rho_{{}_{\rm QES}}, \rho_{{}_{\rm trial}})  \hspace{0.2cm}  $ \\
\hline
\hline
$0$ & $n=0$ & $1.24 \times 10^{-11}$ \\
\hline
\multirow{2}{*}{$1$} & $n=0$ & $2.80 \times 10^{-10}$ \\
\cline{2-2}
 & $n=2$ & $9.39 \times 10^{-11}$ \\
\hline
\multirow{2}{*}{$2$} & $n=0$ & $5.38 \times 10^{-11}$ \\
\cline{2-2}
 & $n=2$ & $5.84 \times 10^{-10}$ \\
 \hline
 \multirow{2}{*}{$3$} & $n=0$ & $5.38 \times 10^{-11}$ \\
\cline{2-2}
 & $n=2$ & $1.58 \times 10^{-9}$ \\
\hline
\end{tabular}
\caption{{\small Comparison between exact QES probability densities $\rho_{{}_{\rm QES}}$ and the approximate variational densities $\rho_{{}_{\rm trial}}$ obtained from (\ref{trialf}) for $\lambda=0,1,2,3$ and lowest even states $n=0,2$.}}
\label{TCRJQes}
\end{table}

\clearpage

\begin{table}[H]
\centering
\renewcommand{\arraystretch}{1.2}
\setlength{\tabcolsep}{5pt}
\scriptsize
\caption{\small Summary of informational and uncertainty measures in the QES sextic potential. }
\vspace{0.3em}
\begin{tabular}{|>{\centering\arraybackslash}p{3.2cm}|
                >{\centering\arraybackslash}p{4.2cm}|
                >{\centering\arraybackslash}p{5cm}|
                >{\centering\arraybackslash}p{2.8cm}|}
\hline
\textbf{Measure} & \textbf{Observed Behavior} & \textbf{Physical Insight} & \textbf{Sensitivity to Transitions} \\
\hline
\( \Delta x \) \newline (Position Uncertainty) &
Monotonic (except $n=2$); inter-pair crossings are observed, but no intra-pair level crossings occurs ; quasi-pairing &
Signals wavefunction broadening and real-space delocalization; ($n=0$, $n=1$) and ($n=2$, $n=3$) form pairs&
Moderate--High \\
\hline
\( \Delta p \) \newline (Momentum Uncertainty) &
Non-monotonic; no level crossing occurs; displays a minimum near \( \lambda_c^{(n)} \); quasi-pairing  &
Tracks compression in momentum space due to delocalization in \( x \) &
Moderate \\
\hline
\( \Delta x \Delta p \) \newline (Heisenberg Product) &
Quasi-pairing of \((n=0,1)\), \((n=2,3)\); inter-pair crossings are observed, but no intra-pair level crossings appears; quasi-pairing&
Global uncertainty indicator; very similar to harmonic oscillator at low \( \lambda \) and lowest states $n=0,1,2,3$&
Moderate \\
\hline
\( S_x \) \newline (Position Shannon Entropy) &
Non-monotonic; local maxima near after \( \lambda_c^{(n)} \); inter- and intra-pair crossings are observed &
Captures delocalization and symmetry breaking beyond variance-based methods &
High \\
\hline
\( S_p \) \newline (Momentum Shannon Entropy) &
Non-monotonic; partially complementary to \( S_x \); intra-pair crossings are observed, but no inter-pair level crossings occurs &
Reveals changes in momentum-space localization &
Moderate \\
\hline
\( S_t = S_x + S_p \) \newline (Entropic Sum) &
Crossings akin to those in $S_p$ are present; odd-$n$th-states less affected by $\lambda$&
Encodes total quantum uncertainty information; delayed onset of entropic pairing at large $\lambda$ &
High \\
\hline
KL Divergence \newline \( D_{KL}(\rho_1 || \rho_0) \) &
Smooth monotonic decreasing function of $\lambda$&
Quantifies shape deviation in $n=1$ from ground state $n=0$ in core regions &
Very High \\
\hline
CRJ Divergence \newline \( D_{CRJ}(\rho_n || \rho_{n+1}) \) &
Non-monotonic for excited pairs like ($n=2,n=3$); vanishes at large $\lambda$&
Sensitive to tail structure and near-degenerate state formation &
Very High \\
\hline
Wavefunction Shape \newline (Real Space) &
Even \( n \): double-peaked; Odd \( n \): less deformed &
Signature of tunneling, parity effects, and symmetry loss &
High \\
\hline
State Pairing (position-space)&
Emerges for \((n=0,1)\), \((n=2,3)\); quasi-degenerate levels; delayed onset of entropic pairing at large $\lambda$ &
Reflects formation of symmetric/antisymmetric doublets &
High \\
\hline
Algebraic Sector Boundary &
Begins at \( \lambda = 0 \); structure changes beyond &
The appearance of QES exact solutions is not directly reflected in the behavior of entropic measures &
Moderate \\
\hline
\end{tabular}
\label{Tsumm}
\end{table}

\section{Conclusion}
\label{s7}

In this work, we have analyzed the quasi-exactly solvable sextic potential with a tunable parameter $\lambda$, which interpolates between a single-well and a symmetric double-well structure. Our investigation combined variational methods, Lagrange mesh numerics, and information-theoretic tools to study the first four quantum states $n = 0, 1, 2, 3$ over a broad range of $\lambda$. 

A key finding is the emergence of harmonic-like behavior in the Heisenberg uncertainty product $\Delta x\,\Delta p$ for these lowest states, particularly below the critical coupling $\lambda_n^c$ where the potential transitions from a single- to a double-well. The nearly constant spacing in $\Delta x\,\Delta p$ mirrors the behavior of the quantum harmonic oscillator. However, while this uncertainty product reveals overall confinement trends, it fails to capture more subtle structural transitions such as localization asymmetry, tunneling onset, and parity-induced interference effects.

To that end, we have shown that information-theoretic measures, including Shannon entropy in position and momentum space, Kullback-Leibler (KL) divergence, and Cumulative Residual Jeffreys (CRJ) divergence, are significantly more sensitive. For example, while $\Delta x\,\Delta p$ remains nearly unchanged as the system crosses from a single to a double-well configuration, both KL and CRJ divergences show sharp peaks or discontinuities, corresponding to symmetry breaking, wavefunction bifurcation, and the onset of tunneling-induced pairing. These measures not only signal the transition but also encode qualitative differences between even and odd states, including differences in central nodal structures and momentum-space delocalization.

Another central contribution of this work lies in its dynamical approach to structural transitions. To the best of the authors' knowledge, this is the first study that analyzes the continuous transition from a single-well to a double-well quasi-exactly solvable potential by varying a parameter across a family of potentials. Unlike most works that fix a specific double-well shape and examine its eigenstates, our approach provides a broader understanding of how spectral and informational properties evolve with the potential's profile.

Finally, we observe that energy levels naturally form nearly degenerate pairs—such as $(n = 0,1)$ and $(n = 2,3)$—for large $\lambda$, a behavior consistent with semiclassical WKB predictions for symmetric double wells. Even and odd states within each pair exhibit distinct behavior: even-parity states develop central dips or near-nodes, while odd-parity states maintain exact nodes at the origin. These differences persist in momentum space and are readily detected by entropic measures but not by Heisenberg uncertainties.

Summarizing, this study provides a comprehensive and nuanced picture of spectral evolution in a quasi-exactly solvable system with tunable topology, combining algebraic solvability, information geometry, and quantum tunneling. 

Moreover, the ability to characterize tunneling, coherence, and localization transitions using entropic and divergence-based diagnostics in quasi-exactly solvable systems provides valuable insights for quantum technologies. Such analyses can inform the design of controllable quantum wells in quantum dots, the engineering of qubit potentials in superconducting circuits, and the development of protocols for quantum state preparation and manipulation in systems where analytical tractability and tunability are essential.

Building on the present informational and divergence-based analysis, future work will incorporate the study of Wigner quasiprobability distributions to visualize the full phase-space structure of the eigenstates across the single- to double-well transition. This approach will provide deeper insight into the interplay between position and momentum localization, quantum interference, and coherence. In parallel, a more refined semiclassical analysis based on the WKB approximation will be pursued to extract analytical estimates of the energy spectrum. These WKB results, together with the available exact and variational data, will serve as the foundation for constructing a closed-form interpolation \( E = E(\lambda) \) for the lowest energy states. It would be particularly interesting to examine the analytic properties and structural behavior of such interpolations in the complex-\( \lambda \) domain, as this could shed light on connections with spectral singularities, PT symmetry, and the broader analytic structure of quasi-exactly solvable models.



\section*{Data availability}
Data sharing is not applicable to this article as no new data were created or analyzed in this study.

\section*{References}

\bibliographystyle{abbrv}
\bibliography{Biblio}
\bibliographystyle{unsrt}

\clearpage

\appendix

\section{Optimal variational parameters}
\label{Ap1}

Here, for the lowest states $n=0,1,2,3$, we present the optimal values variational parameters of the trial function (\ref{trialf}) for different values of $\lambda$.

\begin{table}[ht]
  \centering
  \renewcommand{\arraystretch}{1.2}
  \resizebox{\textwidth}{!}{
  \begin{tabular}{c|cccccccc}
    \hline
    \rowcolor{gray!10} \multicolumn{9}{c}{\textbf{$\lambda = - \frac{3}{4}$}} \\
    \hline
    \textbf{$n$} & \textbf{$a_{1}$} & \textbf{$a_{2}$} & \textbf{$a_{3}$} & \textbf{$a_{4}$} & \textbf{$a_{5}$} &  \textbf{$a_{6}$} &  \textbf{$a_{7}$} &  \textbf{$a_{8}$} \\
    \hline
    0  & -0.87753969 & 0.44456930 & -0.16059741 & 0.03949804 & -0.00566654 & 0.00034654 & 0 & 0  \\
    2  & -6.02648097 & 6.30032119 & -3.80045170 & 1.56945998 & -0.44438189 & 0.08096664 & -0.00841404 & 0.00037468 \\
    \hline
    \hline
    \textbf{$n$} & \textbf{$b_{1}$} & \textbf{$b_{2}$} & \textbf{$b_{3}$} & \textbf{$b_{4}$} & \textbf{$b_{5}$} &  \textbf{$b_{6}$} &  \textbf{$b_{7}$} &  \textbf{$b_{8}$} \\
    \hline
    1  & -1.01467419 & 0.56605256 & -0.20964529 & 0.04987984 & -0.00671684 & 0.00038167 & 0 & 0 \\
    3  & -3.15409437 & 3.19435411 & -1.91604400 & 0.77048822 & -0.20776652 & 0.03559229 & -0.00346190 &  0.00014433 \\
    \hline
  \end{tabular}
  }
  \caption{Case $\lambda= - \frac{3}{4}$. For the lowest states $n=0,1,2,3$ the values of the variational parameters $a_i$ (even states) and $b_i$ (odd states) appearing in the trial function (\ref{trialf}).}
  \label{tab:coeficientes-N-34}
\end{table}

\begin{table}[ht]
  \centering
  \renewcommand{\arraystretch}{1.2}
  \resizebox{\textwidth}{!}{
  \begin{tabular}{c|cccccccc}
    \hline
    \rowcolor{gray!10} \multicolumn{9}{c}{\textbf{$\lambda = - \frac{1}{2}$}} \\
    \hline
    \textbf{$n$} & \textbf{$a_{1}$} & \textbf{$a_{2}$} & \textbf{$a_{3}$} & \textbf{$a_{4}$} & \textbf{$a_{5}$} &  \textbf{$a_{6}$} &  \textbf{$a_{7}$} &  \textbf{$a_{8}$} \\
    \hline
    0  & -0.76247145 & 0.33713508 & -0.10980626 & 0.02534971 & -0.00351266 & 0.00021077  & 0 & 0 \\
    2  & -5.64931168 & 5.50715704 & -3.13205803 & 1.23914373 & -0.34115756 & 0.06108501 & -0.00627954 & 0.00027770 \\
    \hline
    \hline
    \textbf{$n$} & \textbf{$b_{1}$} & \textbf{$b_{2}$} & \textbf{$b_{3}$} & \textbf{$b_{4}$} & \textbf{$b_{5}$} &  \textbf{$b_{6}$} &  \textbf{$b_{7}$} &  \textbf{$b_{8}$} \\
    \hline
    1  & -0.92381450 & 0.47566820 & -0.16644310 & 0.03823671 & -0.00504442 & 0.00028318  & 0 & 0  \\
    3  & -2.97890647 & 2.77434291 & -1.47577880 & 0.49721948 & -0.10349375 & 0.01199314 & -0.00058596 &   0 \\
    \hline
  \end{tabular}
  }
  \caption{Case $\lambda= - \frac{1}{2}$. For the lowest states $n=0,1,2,3$ the values of the variational parameters $a_i$ (even states) and $b_i$ (odd states) appearing in the trial function (\ref{trialf}).}
  \label{tab:coeficientes-N-12}
\end{table}

\begin{table}[ht]
\centering
\renewcommand{\arraystretch}{1.6}
 \resizebox{\textwidth}{!}{
  \begin{tabular}{c|cccccccc}
    \hline
    \rowcolor{gray!10} \multicolumn{9}{c}{\textbf{$\lambda = 0$}} \\
    \hline
    \textbf{$n$} & \textbf{$a_{1}$} & \textbf{$a_{2}$} & \textbf{$a_{3}$} & \textbf{$a_{4}$} & \textbf{$a_{5}$} & \textbf{$a_{6}$} & \textbf{$a_{7}$} &  \textbf{$a_{8}$} \\
    \hline
    0  & -0.49981340 & 0.12432545 & -0.01995370 & 0.00207924 & -0.00010713 & 0 & 0 & 0 \\
    2  & -4.86743892 & 4.00458277 & -1.95992950 & 0.69267964 & -0.17706875 & 0.03029207 & -0.00302875 & 0.00013163 \\
    \hline
    \hline
    \textbf{$n$} & \textbf{$b_{1}$} & \textbf{$b_{2}$} & \textbf{$b_{3}$} & \textbf{$b_{4}$} & \textbf{$b_{5}$} & \textbf{$b_{6}$} & \textbf{$b_{7}$} &  \textbf{$b_{8}$} \\
    \hline
    1  & -0.72501829 & 0.29637523 & -0.08685787 & 0.01782717 & -0.0020360 & 0.00011916 & 0 & 0 \\
    3  & -2.68351516 & 2.22839688 & -1.08149686 & 0.34191319 & -0.06831479 & 0.00771500 & -0.00037077 & 0 \\
    \hline
\end{tabular} 
}
\caption{Case $\lambda=0$. For the lowest states $n=0,1,2,3$ the values of the variational parameters $a_i$ (even states) and $b_i$ (odd states) appearing in the trial function (\ref{trialf}).}
\label{tab:coeficientes-N0} 
\end{table}

\begin{table}[ht]
  \centering
  \renewcommand{\arraystretch}{1.6}
  \resizebox{\textwidth}{!}{
  \begin{tabular}{c|cccccccc}
    \hline
    \rowcolor{gray!10} \multicolumn{9}{c}{\textbf{$\lambda = \frac{1}{2}$}} \\
    \hline
    \textbf{$n$} & \textbf{$a_{1}$} & \textbf{$a_{2}$} & \textbf{$a_{3}$} & \textbf{$a_{4}$} & \textbf{$a_{5}$} & \textbf{$a_{6}$} & \textbf{$a_{7}$} & \textbf{$a_{8}$} \\
    \hline
    0  & -0.18000176 & -0.06910089 & 0.03677595 & -0.00703956 & 0.00052041  & 0 & 0 & 0 \\
    2  & -4.05317006 & 2.63338223 & -0.99605780 & 0.26707226 & -0.04961702 & 0.00554925 & -0.00027324 & 0.00013163 \\
    \hline
    \hline
    \textbf{$n$} & \textbf{$b_{1}$} & \textbf{$b_{2}$} & \textbf{$b_{3}$} & \textbf{$b_{4}$} & \textbf{$b_{5}$} & \textbf{$b_{6}$} & \textbf{$b_{7}$} &  \textbf{$b_{8}$} \\
    \hline
    1  & -0.49684377 & 0.11892573 & -0.01627533 & 0.00102145 & 0 & 0 & 0 & 0 \\
    3  & -2.37320352 & 1.69851842 & -0.72611484 & 0.20953608 & -0.03946687 & 0.00429658 & -0.00020179 & 0  \\
    \hline
  \end{tabular}
  }
  \caption{Case $\lambda=\frac{1}{2}$. For the lowest states $n=0,1,2,3$ the values of the variational parameters $a_i$ (even states) and $b_i$ (odd states) appearing in the trial function (\ref{trialf}).}
  \label{tab:coeficientes-N12}
\end{table}

\begin{table}[ht]
  \centering
  \renewcommand{\arraystretch}{1.6}
  \resizebox{\textwidth}{!}{
  \begin{tabular}{c|ccccccc}
    \hline
    \rowcolor{gray!10} \multicolumn{8}{c}{\textbf{$\lambda_{c}^{n=0} = 0.7329531261$}} \\
    \hline
    \textbf{$n$} & \textbf{$a_{1}$} & \textbf{$a_{2}$} & \textbf{$a_{3}$} & \textbf{$a_{4}$} & \textbf{$a_{5}$} & \textbf{$a_{6}$} &  \textbf{$a_{7}$} \\
    \hline
    0  & -0.00225464 & -0.15246795 & 0.05456054 & -0.00906889 & 0.00062194 & 0 & 0 \\
    2  & -3.67090635 & 2.07582964 & -0.67157652 & 0.15615390 & -0.02615931 & 0.00274346 & -0.00013007   \\
    \hline
    \hline
    \textbf{$n$} & \textbf{$b_{1}$} & \textbf{$b_{2}$} & \textbf{$b_{3}$} & \textbf{$b_{4}$} & \textbf{$b_{5}$} & \textbf{$b_{6}$} &  \textbf{$b_{7}$} \\
    \hline
    1  & -0.38620385 & 0.05358857 & 0.00087360 & -0.00118690 & 0.00010805 & 0 & 0 \\
    3  & -2.22402694 & 1.46026286 & -0.57671313 & 0.15675104 & -0.02838486 & 0.00301593 & -0.00013954 \\
    \hline
  \end{tabular}
  }
  \caption{ Case $\lambda_{c}^{n=0} = 0.7329531261$. For the lowest states $n=0,1,2,3$ the values of the variational parameters $a_i$ (even states) and $b_i$ (odd states) appearing in the trial function (\ref{trialf}).}
  \label{tab:coeficientes-Nc}
\end{table}

\begin{table}[ht]
  \centering
  \renewcommand{\arraystretch}{1.2}
  \resizebox{\textwidth}{!}{
  \begin{tabular}{c|cccccc}
    \hline
    \rowcolor{gray!10} \multicolumn{7}{c}{\textbf{$\lambda = 1.5$}} \\
    \hline
    \textbf{$n$} & \textbf{$a_{1}$} & \textbf{$a_{2}$} & \textbf{$a_{3}$} & \textbf{$a_{4}$} & \textbf{$a_{5}$} &  \textbf{$a_{6}$}  \\
    \hline
    0  & 0.77643421 & -0.32757503 & 0.04028137 & -0.00010013 & -0.00025120 & 0 \\
    2  & -2.42832724 & 0.57949804 & 0.02713633 & -0.03456271 & 0.00631668 & -0.00040744 \\
    \hline
    \hline
    \textbf{$n$} & \textbf{$b_{1}$} & \textbf{$b_{2}$} & \textbf{$b_{3}$} & \textbf{$b_{4}$} & \textbf{$b_{5}$} &  \textbf{$b_{6}$}  \\
    \hline
    1  & 0,04616241 & -0.14345103 & 0.04171957 & -0.00572411 & 0.00032955 & 0 \\
    3  & -1.71425726 & 0.72917128 & -0.16813346 & 0.02480347 & -0.00226565 & 0.00009728 \\
    \hline
  \end{tabular}
  }
  \caption{Case $\lambda=\frac{3}{2}$. For the lowest states $n=0,1,2,3$ the values of the variational parameters $a_i$ (even states) and $b_i$ (odd states) appearing in the trial function (\ref{trialf}).}
  \label{tab:coeficientes-N15}
\end{table}

\begin{table}[ht]
  \centering
  \renewcommand{\arraystretch}{1.2}
  \resizebox{\textwidth}{!}{
  \begin{tabular}{c|cccccc}
    \hline
    \rowcolor{gray!10} \multicolumn{7}{c}{\textbf{$\lambda = 2$}} \\
    \hline
    \textbf{$n$} & \textbf{$a_{1}$} & \textbf{$a_{2}$} & \textbf{$a_{3}$} & \textbf{$a_{4}$} & \textbf{$a_{5}$} &  \textbf{$a_{6}$} \\
    \hline
    0  & 1.50132881 & -0.21425641 & -0.09378895 & 0.03554272 & -0.00505270 & 0.00028335 \\
    2  & -1.67337629 & -0.10873188 & 0.23141337 & -0.06359743 & 0.00822978 & -0.00044290 \\
    \hline
    \hline
    \textbf{$n$} & \textbf{$b_{1}$} & \textbf{$b_{2}$} & \textbf{$b_{3}$} & \textbf{$b_{4}$} & \textbf{$b_{5}$} &  \textbf{$b_{6}$} \\
    \hline
    1  & 0.04616241 & -0.14345103 & 0.04171957 & -0.00572411 & 0.00032955 & 0 \\
    3  & -1.71425726 & 0.72917128 & -0.16813346 & 0.02480347 & -0.00226565 & 0.00009728 \\
    \hline
  \end{tabular}
  }
  \caption{Case $\lambda=2$. For the lowest states $n=0,1,2,3,4,5$ the values of the variational parameters $a_i$ (even states) and $b_i$ (odd states) appearing in the trial function (\ref{trialf}).}
  \label{tab:coeficientes-N2}
\end{table}

\begin{table}[ht]
  \centering
  \renewcommand{\arraystretch}{1.2}
  \resizebox{\textwidth}{!}{
  \begin{tabular}{c|cccccc}
    \hline
    \rowcolor{gray!10} \multicolumn{7}{c}{\textbf{$\lambda = 2.5$}} \\
    \hline
    \textbf{$n$} & \textbf{$a_{1}$} & \textbf{$a_{2}$} & \textbf{$a_{3}$} & \textbf{$a_{4}$} & \textbf{$a_{5}$} &  \textbf{$a_{6}$} \\
    \hline
    0  & 2.447028601 & 0.25176410 & -0.30768497 & 0.06421671 & -0.00611020 & 0.00023715 \\
    2  & -0.98511016 & -0.58885186 & 0.27441005 & -0.04217453 & 0.00244557 & 0 \\
    \hline
    \hline
    \textbf{$n$} & \textbf{$b_{1}$} & \textbf{$b_{2}$} & \textbf{$b_{3}$} & \textbf{$b_{4}$} & \textbf{$b_{5}$} &  \textbf{$b_{6}$} \\
    \hline
    1  & 0.76259748 & -0.19747528 & -0.00597625 & 0.00615133 & -0.00055238 & 0 \\
    3  & -1.03061288 & -0.00330500 & 0.08777061 & -0.01913043 & 0.00136000 & 0 \\
    \hline
  \end{tabular}
  }
  \caption{Case $\lambda=\frac{5}{2}$. For the lowest states $n=0,1,2,3$ the values of the variational parameters $a_i$ (even states) and $b_i$ (odd states) appearing in the trial function (\ref{trialf}).}
  \label{tab:coeficientes-N25}
\end{table}

\begin{table}[ht]
  \centering
  \renewcommand{\arraystretch}{1.2}
  \resizebox{\textwidth}{!}{
  \begin{tabular}{c|cccccc}
    \hline
    \rowcolor{gray!10} \multicolumn{7}{c}{\textbf{$\lambda = 3$}} \\
    \hline
    \textbf{$n$} & \textbf{$a_{1}$} & \textbf{$a_{2}$} & \textbf{$a_{3}$} & \textbf{$a_{4}$} & \textbf{$a_{5}$} &  \textbf{$a_{6}$} \\
    \hline
    0  & 3.61074888 & 1.28848697 & -0.51894892 & 0.04153164 & 0.00316484 & 0.00047766 \\
    2  & -0.33150853 & -0.91546494 & 0.19631170 & 0.00683826 & -0.00555149 & 0.00046010 \\
    \hline
    \hline
    \textbf{$n$} & \textbf{$b_{1}$} & \textbf{$b_{2}$} & \textbf{$b_{3}$} & \textbf{$b_{4}$} & \textbf{$b_{5}$} &  \textbf{$b_{6}$} \\
    \hline
    1  & 1.17769976 & -0.03444628 & -0.09841690 & 0.02432078 & -0.00257698 & 0.00011136 \\
    3  & -0.65878502 & -0.31424786 & 0.15193356 & -0.02340663 & 0.00134675 & 0 \\
    \hline
  \end{tabular}
  }
  \caption{Case $\lambda=3$. For the lowest states $n=0,1,2,3$ the values of the variational parameters $a_i$ (even states) and $b_i$ (odd states) appearing in the trial function (\ref{trialf}).}
  \label{tab:coeficientes-N3}
\end{table}

\begin{table}[ht]
  \centering
  \renewcommand{\arraystretch}{1.2}
  \resizebox{\textwidth}{!}{
  \begin{tabular}{c|ccccccc}
    \hline
    \rowcolor{gray!10} \multicolumn{8}{c}{\textbf{$\lambda = 3.5$}} \\
    \hline
    \textbf{$n$} & \textbf{$a_{1}$} & \textbf{$a_{2}$} & \textbf{$a_{3}$} & \textbf{$a_{4}$} & \textbf{$a_{5}$} &  \textbf{$a_{6}$} &  \textbf{$a_{7}$} \\
    \hline
    0  & 4.98209852 & 3.06058862 & -0.42379567 & -0.15872775 & 0.04981656 & -0.00568678 & 0.00025410 \\
    2  & 0.34491645 & -1.07867772 & -0.03580624 & 0.08838160 & -0.01610316 & 0.00097790 & 0  \\
    \hline
    \hline
    \textbf{$n$} & \textbf{$b_{1}$} & \textbf{$b_{2}$} & \textbf{$b_{3}$} & \textbf{$b_{4}$} & \textbf{$b_{5}$} &  \textbf{$b_{6}$} &  \textbf{$b_{7}$} \\
    \hline
    1  & 1.64492086 & 0.27736894 & -0.18115974 & 0.02463137 & -0.00099738 & -0.00002128 & 0 \\
    3  & -0.27324165 & -0.54402271 & 0.13880345 & -0.00581923 & -0.00148710 & 0.00014703 & 0 \\
    \hline
  \end{tabular}
  }
  \caption{Case $\lambda=3.5$. For the lowest states $n=0,1,2,3$ the values of the variational parameters $a_i$ (even states) and $b_i$ (odd states) appearing in the trial function (\ref{trialf}).}
  \label{tab:coeficientes-N35}
\end{table}

\begin{table}[ht]
  \centering
  \renewcommand{\arraystretch}{1.2}
  \resizebox{\textwidth}{!}{
  \begin{tabular}{c|ccccccc}
    \hline
    \rowcolor{gray!10} \multicolumn{8}{c}{\textbf{$\lambda = 4$}} \\
    \hline
    \textbf{$n$} & \textbf{$a_{1}$} & \textbf{$a_{2}$} & \textbf{$a_{3}$} & \textbf{$a_{4}$} & \textbf{$a_{5}$} &  \textbf{$a_{6}$} &  \textbf{$a_{7}$} \\
    \hline
    0  & 6.43399795 & 6.07255642 & -0.10263771 & -0.28107904 & 0.03021223 & 0 & 0 \\
    2  & 1.12208848 & -1.03009518 & -0.44483765 & 0.17776174 & -0.02096813 & 0.00071626 & 0.00002106 \\
    \hline
    \hline
    \textbf{$n$} & \textbf{$b_{1}$} & \textbf{$b_{2}$} & \textbf{$b_{3}$} & \textbf{$b_{4}$} & \textbf{$b_{5}$} &  \textbf{$b_{6}$} &  \textbf{$b_{7}$} \\
    \hline
    1  & 2.13578290 & 0.82862981 & -0.25720280 & 0.01700864 & 0.00019329 & 0 & 0  \\
    3  & 0.12641765 & -0.66406584 & 0.03412807 & 0.03322343 & -0.00662624 & 0.00040093 & 0 \\
    \hline
  \end{tabular}
  }
  \caption{Case $\lambda=4$. For the lowest states $n=0,1,2,3$ the values of the variational parameters $a_i$ (even states) and $b_i$ (odd states) appearing in the trial function (\ref{trialf}).}
  \label{tab:coeficientes-N4}
\end{table}

\clearpage

\section{WKB analysis in the asymptotic regimes $\lambda \pm \infty$}
\label{Ap2}

We analyze the semiclassical behavior of the quasi-exactly solvable potential
\begin{equation}
V^{\text{QES}}(x; \lambda) = \frac{1}{2}\left(x^6 + 2x^4 - 2(2\lambda + 1)x^2\right),
\end{equation}
in the asymptotic regimes \( \lambda \to +\infty \) and \( \lambda \to -\infty \), using the WKB approximation. The associated semiclassical quantization condition is
\begin{equation}
\int_{x_1}^{x_2} \sqrt{2(E - V(x))} \, dx = \left(n + \frac{1}{2}\right)\pi,
\end{equation}
where \( x_1 \) and \( x_2 \) denote the classical turning points at energy \( E \).

\subsection*{The Limit \( \lambda \to +\infty \)}

To analyze the spectrum in the $\lambda \to \infty$ limit, we first rescale the coordinate as
\begin{equation}
x = \lambda^{1/4} y,
\end{equation}
which transforms the potential into
\begin{equation}
V(x) = \lambda^{3/2} \cdot \frac{1}{2} \left( y^6 + 2\lambda^{-1/2} y^4 - 4\lambda^{-1} y^2 \right).
\end{equation}
To leading order in $\lambda$, this simplifies to
\begin{equation}
V(x) \approx \lambda^{3/2} \cdot \frac{1}{2}(y^6 - 4y^2) \equiv \lambda^{3/2} U(y),
\end{equation}
where $U(y) = \frac{1}{2}(y^6 - 4y^2)$ defines the rescaled potential.

The total energy accordingly rescales as
\begin{equation}
E = \lambda^{3/2} \epsilon.
\end{equation}
In the semiclassical (WKB) framework, the quantization condition in the rescaled coordinate becomes
\begin{equation}
\int_{y_1}^{y_2} \sqrt{2\epsilon - (y^6 - 4y^2)}\, dy = \left(n + \frac{1}{2}\right)\pi \cdot \lambda^{-3/4},
\end{equation}
where $y_1$ and $y_2$ are classical turning points satisfying
\begin{equation}
y^6 - 4y^2 = 2\epsilon.
\end{equation}
For the integral to be real, the integrand must be non-negative, i.e.,
\begin{equation}
2\epsilon \ge y^6 - 4y^2.
\end{equation}
Since the function $y^6 - 4y^2$ is negative in a finite region around $y = \pm\sqrt{2}$, the condition above implies
\begin{equation}
\epsilon_n < 0.
\end{equation}
Thus, all low-lying semiclassical states are localized in the classically allowed region where $U(y) < 0$. Their physical energies scale as
\begin{equation}
E_n = \lambda^{3/2} \epsilon_n, \quad \text{with } \epsilon_n < 0,
\end{equation}
implying that
\begin{equation}
E_n \to -\infty \quad \text{as} \quad \lambda \to +\infty.
\end{equation}

Owing to the symmetry of the potential $U(y)$, the physical eigenstates are either symmetric or antisymmetric combinations of wavefunctions localized in the left and right wells. While the WKB quantization accurately captures the bound-state energies of an individual well, the true eigenstates exhibit a small energy splitting due to quantum tunneling through the central barrier:
\begin{equation}
E_n^{\pm} = E_n^0 \pm \delta E_n,
\end{equation}
where $E_n^0 = \lambda^{3/2} \epsilon_n$ is the energy from single-well quantization, and $\delta E_n$ represents the tunneling-induced splitting. This splitting is exponentially suppressed in the semiclassical limit and may be estimated using the WKB action across the classically forbidden region:
\begin{equation}
\delta E_n \sim e^{-S_n/\hbar}, \quad
S_n = \int_{y_-}^{y_+} \sqrt{y^6 - 4y^2 - 2\epsilon_n} \, dy,
\end{equation}
where $y_\pm$ are the classical turning points under the barrier. This exponential suppression encodes the parity structure of the spectrum and is a hallmark of symmetric double-well potentials.

\subsection{The Limit $\lambda \to -\infty$}

For large negative $\lambda$, the potential becomes a steep, single-well centered at the origin. No rescaling is required, and the leading contribution is:
\begin{equation}
V(x) \ \approx \  2\,|\lambda|\, x^2,
\end{equation}
which corresponds to a harmonic oscillator with frequency $\omega = \sqrt{4|\lambda|}$. The WKB spectrum is thus:
\begin{equation}
E_n(\lambda) \ \approx \ \sqrt{|\lambda|} \left(1\, +\,2\,n \,\right), \quad \lambda \to -\infty.
\end{equation}

Hence, in this limit the eigenvalues grow as $|\lambda|^{1/2}$, in contrast to the $\lambda^{3/2}$ growth observed in the positive regime.

This analysis reveals a sharp transition in the qualitative nature of the spectrum, from harmonic-like at large negative $\lambda$, to deeply double-well and semiclassically degenerate at large positive $\lambda$. 

\end{document}